\documentclass[aps,prd,preprint,amsmath,citeautoscript,longbibliography,nofootinbib]{revtex4-2} 
\usepackage[utf8]{inputenc}
\usepackage{bm}
\usepackage{amsmath,mathrsfs,amsfonts}
\usepackage{graphicx,afterpage}
\usepackage{subcaption}
\usepackage{amsmath,latexsym}
\usepackage{color,soul}
\usepackage{float}
\usepackage[section]{placeins}
\usepackage{silence}
\begin{document}
\title{NUT Solutions in Einstein-Maxwell-scalar-Gauss-Bonnet Gravity}
\author{M. Butler{\footnote{m.butler@usask.ca}} and A. M. Ghezelbash{\footnote{masoud.ghezelbash@usask.ca}}}

\affiliation{Department of Physics and Engineering Physics, University of Saskatchewan, Saskatoon SK S7N 5E2, Canada}
\date{\today}

\begin{abstract}
In this article, we consider a class of  four-dimensional Einstein-Maxwell theory which is coupled non-minimally to a scalar field and the Gauss-Bonnet invariant. We mainly use the numerical methods to find the solutions to the theory, with the NUT twist.
We find explicitly the numerical solutions to all of the field equations. To find the appropriate consistent numerical solutions, we use the perturbative expansion of the fields asymptotically, as well as near horizon. The solutions describe the NUTty black holes with the scalar charge which also depend explicitly on the values of the non-minimal coupling constants. We provide a detailed numerical analysis of the solutions in terms of all existing parameters of the theory.
\end{abstract}

\maketitle

\section{Introduction}
\label{sec:intro}

One of the most interesting class of exact solutions to the Einstein equations are the gravitational instantons. These solutions are regular and complete everywhere and posses a self dual curvature two form. Such solutions exist in in vacuum \cite{1} or with a cosmological constant term \cite{2}.  Among these solutions, Taub-NUT solutions and its bolt extension, are the most well known solutions \cite{3} with a lot of applications such as in higher-dimensional gravity \cite{31}, M-theory \cite{32}, black hole holography \cite{33} and black hole thermodynamics \cite{34}. Other interesting examples of the gravitational instantons are 
Eguchi-Hanson  \cite{4} and Atiyah-Hitchin spaces \cite{5}.  
These solutions play an important role in construction of higher-dimensional solutions to extended theories of gravity  \cite{7}, \cite{neww}, \cite{neww2}, supergravity \cite{8}, \cite{9} and in investigating the quantum properties of the black holes \cite{10}. 

Moreover, in the path-integral formulation of the Euclidean quantum gravity, we encounter the gravitational instantons as the dominant metrics. These dominant metrics are also related to the minimal surfaces in Euclidean space \cite{13}.  In fact, the two dimensional minimal surfaces provide solutions to the real elliptic Monge-Amp\`ere equation on a real two dimensional  manifold. The solutions are the base for the Kahler metrics for some gravitational instantons beside the aforementioned well known solutions \cite{meK}.

Inspired by the existence of the gravitational instantons in four dimensions, in this article, we numerically construct and study a new class of exact NUTty solutions in Einstein-Maxwell-scalar-Gauss-Bonnet (EMsGB) theory. 

We organize the article as follows. In section \ref{sec:TN}, we review briefly the Nutty gravitational instantons and the EMsGB gravity. In section \ref{sec:EOM-Der} we review our procedure for calculating the equations of motion for the Einstein, Maxwell and scalar equations. Next, in section \ref{sec:method} we detail the process of reducing our system coupled 1st and 2nd order differential equations and representing them as 4 coupled 1st order equations. Using the analysis and equations from these sections, we numerically calculate and present our findings in section \ref{sec:results}. We close the article by offering our concluding remarks in section \ref{sec:conc}.

\section{Taub-NUT-de Sitter (TNdS) geometry and Einstein-Maxwell-scalar-Gauss-Bonnet theory}\label{sec:TN}
 
\subsection{Taub-NUT-de Sitter Geometry}
The Euclidean de-Sitter (dS) space with the cosmological constant $\Lambda=3/\ell^2$ and with the NUT twist, are given by the metrics  
{\textcolor{black}{
\begin{equation}
ds^{2} = \xi(\tau)({\textcolor{black}{d\psi-}}2n\cos\chi d\phi )^{2} {\textcolor{black}+} \frac{d\tau ^{2}}{\xi(\tau )}+(\tau
^{2}{\textcolor{black}-}n^{2})(d\chi ^{2}+\sin ^{2}\chi d\phi ^{2}),  \label{eqn:TBDS}
\end{equation}%
where 
\begin{equation}
\xi(\tau )=\frac{\tau ^{4}-\left( 6n^{2}+\ell ^{2}\right) \tau ^{2}+2m\ell
^{2}\tau -n^{2}\left( 3n^{2}+\ell ^{2}\right) }{\left( n^2-\tau
^{2}\right) \ell ^{2}} , \label{eqn:VV}
\end{equation}%
}}
In \eqref{eqn:TBDS} and \eqref{eqn:VV},  $n$  denotes the non-zero NUT charge, which can be regarded as a magnetic
type of mass. The metric \eqref{eqn:TBDS} is an 
exact solution to the Einstein equations, where
the coordinate ${\textcolor{black}{\psi}}$ parametrizes a circle fibered over the
sphere of radius $\tau^2-n^2$, which is parametrized by the coordinates $\chi$ and $\phi$, respectively. The coordinate ${\textcolor{black}{\psi}}$  must have
periodicity$\left| \frac{8\pi n}{k}\right| $ to avoid  any conical singularities,
where $k$ is a positive integer. 
The geometry of a constant-$\tau $ surface describes
a Hopf fibration of $S^{1}$ over $S^{2}$. The metric (\ref%
{eqn:TBDS}) actually represents the contraction or expansion of three dimensional fibered sphere in
parts of the spacetime where $\xi(\tau )>0$ and are outside of the past or future cosmological
horizons. We note that there are not any closed timelike curves in these regions, and so the
metric (\ref{eqn:TBDS})  is asymptotically locally dS, with $\left| k\right| $
points identified around the $S^{1}$ fibres. 
We also note that the spacelike Killing vector $\partial /\partial t$\ has a fixed point set
where $\xi(\tau _{0})=0$. 
The topology of the fixed point set corresponds to a $S^2$. 
Moreover, we note that $\frac{%
\partial }{\partial \phi }$\ is a Killing vector. Hence  for any constant $\phi $-slice near the horizon $\tau =\tau _{0}$, additional conical singularities appear in the $({\textcolor{black}{\psi}},\tau )$\ Euclidean section. These conical singularities disappear if  ${\textcolor{black}{\psi}}$ has
period $4\pi /\left| \xi^{\prime }\left( \tau _{0}\right) \right| $. The
equality of this period with $\left| \frac{8\pi n}{k}\right| $, forces $\tau
_{0}=\tau ^{\pm }$\ where,
\begin{equation}
\tau ^{\pm }=\frac{k\ell ^{2}\pm \sqrt{k^{2}\ell ^{4}-144n^{4}+48n^{2}\ell
^{2}}}{12n} . \label{eqn:taus}
\end{equation}%
Hence we find two different extensions\ (TB$^{\pm }$) to dS spacetime, which we can call
Taub-bolt-de Sitter space. The term ``bolt''   refers to the dimensionality of the fixed-point
set of $\partial /\partial \psi$, which is 2. The mass parameters for the TB$^{\pm }$ are given by%
\begin{equation}
m^{\pm }=-\frac{k^{3}\ell ^{6}\pm (288n^{4}-24\ell ^{2}n^{2}+k^{2}\ell ^{4})%
\sqrt{k^{2}\ell ^{4}-144n^{4}+48n^{2}\ell ^{2}}}{864n^{3}\ell ^{2}}.
\label{eqn:massesonn}
\end{equation}%
We also find that to have real values for $\tau ^{\pm }$ and $m^{\pm}$, the NUT parameter should satisfy
\begin{equation}
\left| n\right| \leq \frac{1}{6}\ell \sqrt{6+3\sqrt{4+k^{2}}} . \label{eqn:nmax}
\end{equation}%
Without loss of generality we can take 
$n>0$; results for $n<0$ can be obtained by reversing the signs of $t$ and $%
\phi $. We should also note that TB$^{-}$\ does not exist for $\left| n\right|
<n_{c}=.2658\,\ell $\ since the function $\xi^{-}(\tau )$ gets two additional
larger real roots. This hinders the periodicity condition.

The metric \eqref{eqn:TBDS} is free of any type of curvature singularities. In fact, we find that  the
invariants $R_{\mu \nu \rho \lambda }R^{\mu \nu \rho \lambda }$\ and $\sqrt{%
-g}R_{\mu \nu }^{\quad \alpha \beta }\epsilon _{\alpha \beta \rho \lambda
}R^{\mu \nu \rho \lambda }$ are finite everywhere. However NUT-charged
spacetimes in general have quasiregular singularities. These singularities are the end points of
incomplete and inextendable geodesics. On these geodesics the Riemann tensor and its
derivatives remain finite, in all parallelly propagated orthonormal frames of references
\cite{Kon}. Moreover, TB$^{\pm }$ have the cosmological singularities. In fact, a well-known  conjecture \cite{bala} states that the scalar Riemann curvature invariants 
diverge, to form timelike regions of geodesic incompleteness, whenever the
conserved mass of the spacetime becomes positive. This means the mass of spacetime becomes  larger than the zero
mass for pure dS. In what follows in this article, we consider the Lorentzian signature of geometry \eqref{eqn:TBDS} without any cosmological constant and try to find proper and regular solutions for the metric functions in the context of EMsGB theory. To our knowledge this article is the first which shows the existence of Nutty solutions in the EMsGB theory. {\textcolor{black}{The Lorentzian signature of geometry \eqref{eqn:TBDS} can be obtained by performing a double analytical continuation of the NUT charge and coordinates $\psi$, as given by $n\rightarrow -iN,\psi \rightarrow it$. Hence in the absence of any cosmological constant, we find the Lorentzian signature of the geometry \eqref{eqn:TBDS}, as
\begin{equation}
ds^{2} = -\xi_0(r)({\textcolor{black}{dt+}}2N\cos\theta d\phi )^{2} {\textcolor{black}+} \frac{dr ^{2}}{\xi_0(r )}+(r
^{2}{\textcolor{black}+}N^{2})(d\theta ^{2}+\sin ^{2}\theta d\phi ^{2}),  \label{eqn:TBDS0}
\end{equation}
where we renamed the coordinates $\tau$ and $\chi$ in \eqref{eqn:TBDS}, to $r$ and $\theta$, respectively. We note that $\xi_0(r)$ in \eqref{eqn:TBDS0} is given by  $\xi_0(r)=\frac{r^2-2mr-N^2}{r^2+N^2}$. 
}}
\subsection{Einstein-Maxwell-scalar-Gauss-Bonnet Theory}
We consider an interesting class of EMsGB theory in four dimensions, where the action is given by \cite{Hunter}
\begin{equation}
S=\frac{1}{16\pi}\int d^4x \sqrt{-g}\{\tfrac{R}{2}-\tfrac{1}{8}F_{\mu\nu}F^{\mu\nu}-\tfrac{1}{2}\partial _\mu \phi \partial ^\mu \phi+\alpha  \phi{\cal G}-\beta \phi F_{\mu\nu}F^{\mu\nu}\}.
\label{eqn:act}
\end{equation}
Here $R$ is the Ricci scalar, $F_{\mu\nu}$ is the Maxwell or field strength tensor, $\phi$ is the scalar function, ${\cal G}$ is the Gauss-Bonnet invariant in four dimensions, and $\alpha$ and $\beta$ are the coupling constants for scalar-Gauss-Bonnet and scalar-Maxwell coupling, respectively.  {\textcolor{black}{In equation \eqref{eqn:act} and through the article,  we use the geometrized units such that $G=c=1$}.
The equations of motions are given as,
\begin{eqnarray}
\Box \phi +\alpha {\cal G}-\beta F_{\mu\nu}F^{\mu\nu} &=&0,\label{eqn:FE-scalar}\\
G_{\mu\nu} - T_{\mu\nu} &=& 0, \label{eqn:FE-GR}\\
\partial _\nu \{ \sqrt{-g}(\tfrac{1}{8}+\beta\phi)F^{\mu\nu}\} &=& 0, \label{eqn:FE-MW}
\end{eqnarray}
where $G_{\mu\nu}$ are the Einstein equations, and $T_{\mu\nu}$ is the energy-momentum tensor which has the following form,
\begin{eqnarray}
T_{\mu\nu}&=&\partial_\mu \phi\partial^\mu \phi-\tfrac{1}{2}g_{\mu\nu}\partial_\lambda \phi \partial^\lambda \phi-\alpha\{g_{\lambda\mu}g_{\eta\nu}+g_{\lambda\nu}g_{\eta\mu}\}\epsilon^{\alpha\beta\gamma\eta}\epsilon^{\zeta\kappa\lambda\psi}R_{\zeta\kappa\alpha\beta}\nabla_{\psi}\partial _\gamma \phi \\
&+& \{4\beta\phi+\tfrac{1}{2}\}F_{\mu\lambda}F_\nu^\lambda\nonumber
-\{\beta\phi+\tfrac{1}{8}\}g_{\mu\nu}F_{\lambda\zeta}F^{\lambda\zeta}.
\end{eqnarray}
The Gauss-Bonnet invariant ${\cal G}$ in \eqref{eqn:act} is given by the quadratic expression
\begin{equation}
{\cal G}=R_{\alpha\beta\gamma\delta}R^{\alpha\beta\gamma\delta}-4R_{\alpha\beta}R_{\alpha\beta}+R^2,
\label{eqn:GB-inv}
\end{equation}
in terms of the Ricci tensor $R_{\alpha\beta}$ and Riemann tensor $R_{\alpha\beta\gamma\delta}$.

{\textcolor{black}{Inspired with the exact NUT solutions \eqref{eqn:TBDS0} in General Relativity, we then consider a similar form to the metric \eqref{eqn:TBDS0} in EMsGB theory, to construct the NUT solutions in EMsGB, as}}
\begin{equation}
ds^2=-e^{A(r)}(dt+2N\cos\theta d\varphi)^2+e^{B(r)}dr^2+(N^2+r^2)(d\theta^2+\sin^2\theta d\varphi^2),
\label{eqn:metric-anz}
\end{equation}
where $N>0$ is the NUT charge and $A(r)$ and $B(r)$ are the unknown radial metric functions. 
With the field equations \eqref{eqn:FE-scalar} - \eqref{eqn:FE-MW} and the metric \eqref{eqn:metric-anz} above, we can calculate the individual equations for this system. All analytic calculations are done with geometrized units such that $G=c=1$. 
Next, we assume a time-independent radially symmetric ansatz for the scalar field with the form,
\begin{equation} 
	\phi = \phi(r).
	\label{eqn:ScalarAnsatz}
\end{equation}
Additionally, to calculate $F^{\mu\nu}$ we consider an ansatz for the 4-potential whose non-zero terms are given as,
\begin{align} 
	\begin{split}
		{\cal A}_t &= {\mathrm V}(r) \, dt, \\
		{\cal A}_\varphi  &= {\mathrm V}(r) (2\, N \cos\theta) \, d\varphi.
	\end{split}
    \label{eqn:MWAnsatz}
\end{align}
Where ${\cal A}_t$, ${\cal A}_\varphi$ are the time and $\varphi$ angular components of the 4-potential. $V(r)$ is the electric potential function. By definition we know that $V'(r)$ is the electric field $E(r)$. Differing from prior research \cite{Hunter}, ${\cal A}$ includes a second non-zero term. This is due to the magnetic mass from the metric \eqref{eqn:metric-anz} directly contributing to the 4-potential. This results from the $2N \cos\theta d\varphi$ term which defines a gravitomagnetic monopole for the spacetime \cite{Gravito-EM}.

\section{Construction of Field Equations}\label{sec:EOM-Der}

\subsection{Scalar Field}

To calculate an expression for the scalar field \eqref{eqn:FE-scalar} we first need to find ${\cal G}$, $\Box\phi$, and $F_{\mu\nu}F^{\mu\nu}$ in the context of the metric \eqref{eqn:metric-anz}. Starting with the the Gauss-Bonnet invariant we find,
\begin{align}
\begin{split}
	{\cal G} &= \frac{1}{2 \left(N^{2}+r^{2}\right)^{4} \left({\rm e}^{B}\right)^{2}}
	\bigg[
	6 \left( 4 r^{2} - \left(3 {\rm e}^{A} N^{2} + 4 N^{2} + 4 r^{2}\right) {\rm e}^{B} \right)\left(N^{2}+r^{2}\right)^{2} A'' \\
	&+ 9 \left(4 r^{2} - \left(9{\rm e}^{A} N^{2} + 4 N^{2} + 4 r^{2}\right) {\rm e}^{B} \right) \left(N^{2}+r^{2}\right)^{2} \left(A' \right)^{2} \\
	&+ \left\{ \left(\left(3{\rm e}^{A} N^{2}+4 N^{2} + 4 r^{2}\right) {\rm e}^{B} -  12 r^{2}\right) \left(N^{2}+r^{2}\right) B' 
	+ 8 N^{2} \left(3{\rm e}^{A} {\rm e}^{B} + 2\right) r \right\} \left(N^{2}+r^{2}\right) A' \\
	&- 4 N^{2} {\rm e}^{A} {\rm e}^{B} \left(\left(N^{2} r +r^{3}\right) B' - 2 N^{2}+6 r^{2}\right)
	\bigg].
\label{eqn:GB-TN}
\end{split}
\end{align}
Then for the d'Alembertian of the scalar field $\phi$ we have,
\begin{equation}
\Box\phi = \frac{
	(N^2 + r^2)\phi'' + \tfrac12 \left\{(N^2 + r^2) A' - (N^2 + r^2)B' + 4r\right\}\phi'}
{\left(N^2 + r^2\right){\rm e}^{B}}.
\label{eqn:SC-DA}
\end{equation}
Finally, with the definition $F_{\mu\nu} = {\cal A}_{\nu , \mu} - {\cal A}_{\mu , \nu}$ and the ansatz in \eqref{eqn:MWAnsatz} we find that the square of the field strength tensor is, 
\begin{equation}
	F_{\mu\nu}F^{\mu\nu} =
	\frac{8 N^2 V^2 }{(N^2 + r^2)^2} 
	-\frac{2 E^2}{{\rm e}^{A + B}},
	\label{eqn:FST2}
\end{equation}
which in the $N=0$ case reproduces exactly the results of \cite{Hunter}. Combining \eqref{eqn:GB-TN}, \eqref{eqn:SC-DA}, \eqref{eqn:FST2} with \eqref{eqn:FE-scalar} produces our first differential equation for the scalar field. The full expression is given in Appendix \ref{App:A} as equation \eqref{eqn:DE-Scalar}.

\subsection{Electric Field and Potential Functions}

Next we find the expressions for the electric field $E(r)$ and the electric potential function $V(r)$. By integrating \eqref{eqn:FE-MW}  we lose $\partial_\nu$ and need to consider the integration of the implicit $J^{\mu}$ on the RHS. Integration of $J^{t}$ is by definition total charge $Q$ of the black hole. From the integration of \eqref{eqn:FE-MW} we define the remaining term on the LHS as ${\cal M}^{\mu\nu} = \sqrt{-g}(\tfrac{1}{8}+\beta\phi)F^{\mu\nu}$. Calculating the individual components of this expression results in 3 antisymmetric pairs, ${\cal M}^{tr}=-{\cal M}^{rt}$, ${\cal M}^{\varphi\theta} = -{\cal M}^{\theta\varphi}$ and ${\cal M}^{\theta t} = -{\cal M}^{t \theta}$. The $(t, r)$ terms are functions of $V'(r)$, where we note again that $E(r)=V'(r)$ by definition. The remaining equations are functions of $V(r)$ only. The $(\varphi, \theta)$ terms are linearly dependent on the $(\theta, t)$ equations, differing by a factor of $2N \cos\theta$. Looking at only the linearly independent equations gives us two unique equations that we use to calculate tentative expressions for $E(r)$ and $V(r)$. For the electric field this gives,
\begin{equation}
	E(r) = \frac{{\rm e}^{(A + B)/2}}{(N^2 + r^2)} \frac{Q}{\left(\beta \phi + \tfrac18\right)}.
	\label{eqn:EF}
\end{equation}
From \eqref{eqn:EF}, $E(r)$ takes on the expected form, being proportional with $\tfrac{1}{r^2}$ as we would expect for static solutions of the electric field. In the case that $N=0$, we recover the electric field found previously \cite{Hunter}.

Similarly we find an expression for the electric potential function,
\begin{equation}
	V(r) = - \frac{\textcolor{black}{N}}{\sqrt{{\rm e}^{A+B}}(N^2 + r^2)} \frac{Q}{\left(\beta \phi + \tfrac18\right)},
	\label{eqn:EP}
\end{equation}
which is only calculable when $N \neq 0$ as the differential equations containing $V(r)$ disappear otherwise. 
Looking at \eqref{eqn:EF}, \eqref{eqn:EP} we find a partial limiting case put on $\beta$, $\phi$. As neither equation can be singular, this requires that $\beta \phi > -\tfrac18$ or $\beta \phi < -\tfrac18$.
This produces two cases, if we assume $\beta \in \mathbb{R}^+$ then $\phi > -\tfrac{1}{8 \beta}$. Otherwise, assuming $\beta \in \mathbb{R}^-$ in the second case then $\phi < -\tfrac{1}{8 \beta}$ gives a negative scalar field. The solutions we present will assume a positive scalar field.

\subsection{Einstein Field Equations}

Next we calculate the Einstein field equations, using our metric ansatz \eqref{eqn:metric-anz} and combined with the Einstein equations \eqref{eqn:FE-GR}. We find 5 non-zero differential equations, $G_{tt}$, $G_{rr}$, $G_{\theta\theta}$, $G_{\varphi\varphi}$, and $G_{t \varphi}=-G_{\varphi t}$. Explicit forms are reported in Appendix \ref{App:A} as equations \eqref{eqn:DE-Gtt} - \eqref{eqn:DE-Gtp}. We find that $G_{\varphi\varphi}$ is a linear combination of $G_{tt}$ and $G_{\theta\theta}$, with $G_{\varphi\varphi}=G_{\theta\theta}$ if $\theta=0$ and $G_{\varphi\varphi}=4N^2 G_{tt}$ if $\theta=\pi$. Similarly, $G_{t\varphi} = C(\theta) G_{\varphi\varphi}$ where $C = (-2N, 2N)$ for $\theta = (\pi, 0)$ respectively. The Einstein equation are then reduced to 3 linearly independent equations $G_{tt}$, $G_{rr}$ and $G_{\theta\theta}$ which we can use when finding our solutions. 

Beginning with $G_{rr}$, we note that as with other previous cases \cite{Hunter}, \cite{Antoniou}, \cite{Bakopoulos}, \cite{Sotiriou} we can express \eqref{eqn:DE-Grr} in the form of a quadratic of ${\rm e}^{B(r)}$ which gives,
\begin{align}
\begin{split}
	0 =&\left( {{\rm e}^{B}} \right) ^{2}
	+ \Biggl[\Bigl[ 15\, \left( \beta\,\phi + \frac{2}{15} \right) E(r)^{2} \left( {N}^{2}+{r}^{2} \right) ^{3}{{\rm e}^{-A}}-2\, \left( {N}^{2}+{r}^{2} \right)^{3} \left( \phi' \right)^{2} \\
	&+ 48\, \left(  \left( {N}^{2}{{\rm e}^{A}} + \frac13 \,{N}^{2} + \frac13 \,{r}^{2} \right) \left( {N}^{2}+{r}^{2} \right) A' - \frac43 \,{N}^{2}r{{\rm e}^{A}} \right) \alpha\, \phi' \\
	&+ 4\, \left(  \left( {N}^{2}+{r}^{2} \right) A' + r \right)  \left( {N}^{2}+{r}^{2} \right) r \Bigr] \\
	&	/ \left({ \left( 4\,{N}^{2}+4\,{r}^{2} \right)  \left( {N}^{2} M^2 \phi \beta-{N}^{2}{{\rm e}^{A}}-{N}^{2}-{r}^{2} \right) }\right)\Biggr] {{\rm e}^{B}} \\
	&- 12\,{\frac {\phi' \, A' \alpha\,{r}^{2}}{{N}^{2} M^2 \phi \beta - {N}^{2}{{\rm e}^{A}}-{N}^{2}-{r}^{2}}}.
	\label{eqn:de-qd-eb}
\end{split}
\end{align}
This allows us to construct a solution for ${\rm e}^{B(r)}$ using the typical quadratic equation,
\begin{equation}
	{\rm e}^B = \frac{- \Lambda \pm \sqrt{\Lambda^2 - 4 \Delta}}{2},
	\label{eqn:qd-eB}
\end{equation}
where by inspection we identify $\Lambda$ and $\Delta$ as,
\begin{align}
\begin{split}
	\Lambda &= \bigg[ 
	\left(N^{2}+r^{2}\right)^{3}  \left(\tfrac{8 E^{2}}{{\rm e}^{A}} \left(\beta  \phi + \tfrac18 \right) 
	- 2 \phi'^{2} \right) \\
	&+ 16 \alpha  \left\{\left(3 {\rm e}^{A} N^{2}+N^{2}+r^{2}\right) \left(N^{2}+r^{2}\right) A' -4 N^{2} r \,{\rm e}^{A}\right\} \phi' \\
	& +4 r \left\{\left(N^{2}+r^{2}\right) A' +r \right\} \left(N^{2}+r^{2}\right) \bigg] \\
	&\big/
	4 \left(N^{2}+r^{2}\right) \left(8 N^{2} \beta  \,{\mathrm V}^{2} \phi +N^{2} {\mathrm V}^{2}-{\rm e}^{A} N^{2}-N^{2}-r^{2}\right),
	\label{eqn:lambda}
\end{split}
\end{align}
\begin{align}
\begin{split}
	\Delta &= \frac{12 A' \phi' \alpha  \,r^{2}}{{\rm e}^{A} N^{2} - 8 N^{2} \left(\beta  \phi + \tfrac18 \right) {\mathrm V}^{2} + N^{2} + r^{2}}.
	\label{eqn:delta}
\end{split}
\end{align}

\noindent In the case that $N=0$, we recover the expression for ${\rm e}^{B(r)}$ found previously \cite{Hunter}. Further, if the Maxwell field is removed by setting $Q$, $\beta=0$ we find exactly the results of \cite{Antoniou}, \cite{Bakopoulos}. Together these serve as a useful check for derivations.

We may also use our result in \eqref{eqn:qd-eB} to construct an expression for $B'$ from the definition of the derivative for an exponential function, which in terms of our quadratic coefficients \eqref{eqn:lambda} \eqref{eqn:delta} is given by,
\begin{equation}
	B' = - \frac{\Delta' + \Lambda' {\rm e}^{B}}{2 {\rm e}^{2B} + \Lambda {\rm e}^{B}}.
	\label{eqn:dB}
\end{equation}
With expressions for ${\rm e}^{B}$ and $B'$ we can then substitute these into $G_{tt}$ and $G_{\theta\theta}$ which are now only in terms of two functions, $A(r)$ and $\phi(r)$. As a final step, we can rearrange and isolate the 2nd order derivatives in $G_{tt}$ and $G_{\theta\theta}$ to form two coupled 2nd order ODEs for $A''$ and $\phi''$. 
\begin{align}
	A'' &= F(r, \phi', \phi, A', A; \alpha, \beta, Q, N), \label{eq:DE-A2}\\
	\phi'' &= G(r, \phi', \phi, A', A; \alpha, \beta, Q, N). \label{eq:DE-P2}
\end{align}
Equations \eqref{eq:DE-A2} - \eqref{eq:DE-P2} do not conform to any known general solutions and due to the size and complexity of $F$ and $G$ other analytical methods are not tenable. Therefore we take a numerical approach to solve these equations. Using a standard reduction of order via substitution we get a system of four 1st order ODEs in the general form ${\bf y}'(x) = {\bf f}(x)$, which permits a numerical approach.

\section{Method}
\label{sec:method}

To solve this problem numerically, we need to characterize the limiting behaviour of our equations on both boundaries. Namely, we need to understand how our functions behave near the horizon and outwards to infinity. We define $r_{H}$ to be the non-zero horizon radius. Then we are interested in the two cases where $r \rightarrow r_{H}$, and as $r \rightarrow \infty$. For each case, we find approximations for $A'$, $A$, $\phi'$, and $\phi$ in the respective limit. We expect our boundaries to behave in a similar manner to Taub-NUT solutions in the context of the Einstein-Maxwell equations. 

\subsection{Near Horizon Limit}

Starting with the near horizon limit, 
{\textcolor{black}
for problems of this kind there are typically two approaches used to explore the limiting behaviour of unknown solutions near the horizon. The first involves doing series expansions of the metric functions directly, in our case that requires expansions of the coupled functions $A(r)$, $B(r)$, $\phi(r)$, $V(r)$, and $E(r)$. This is the procedure used by \cite{ref1}, \cite{Kanti}, \cite{Kanti2} for a similar class of problems. Given the size and complexity of our original system of equations \eqref{eqn:DE-Gtt} - \eqref{eqn:DE-Scalar}, including its reduced form in \eqref{eq:DE-A2}-\eqref{eq:DE-P2}, a Taylor series expansion was found to be intractable with symbolic software. 

The second approach available to us is to examine the near horizon behaviour of a single function, and use that analysis to characterize the behaviour of the other functions as was done in \cite{Antoniou}, \cite{Antoniou2}, \cite{Sotiriou2}, \cite{Hunter}.
Using the second approach, we begin by looking at the behaviour of $A$ near the horizon. We can use $A'$ to characterize the behaviour of $e^{B(r)}$ which we have from \eqref{eqn:de-qd-eb}. Then we will employ the same approach again to conduct our analysis of $A''$ and $\phi''$ in the near horizon limit.
}
We assume a horizon radius $r_H$ that defines the black hole horizon for our problem.  As the solutions approach the horizon limit they must diverge, such that $\tfrac{1}{{\rm e}^{A(r)}} \rightarrow \infty$, $A(r) \rightarrow -\infty$, and $A'(r) \rightarrow \infty$ near the horizon.
We can approximate ${\rm e}^{A(r)}$ in the near horizon limit in terms of $A'$ by doing a Taylor series expansion for $({\rm e}^A)'$ around $(r - r_H)$ and applying the horizon limit,
\begin{equation}
	A' = \frac{({\rm e}^{A})'}{{\rm e}^{A}} \rightarrow \lim_{r \rightarrow r_H} \frac{({\rm e}^{A})'}{{\rm e}^{A}} \approx \frac{a_1}{{\rm e}^{A}}.
	\label{eqn:NH-A1}
\end{equation}
The only remaining term from the expansion of $({\rm e}^A)'$ in \eqref{eqn:NH-A1} is the 1st series coefficient $a_1$, and by rearranging we have ${\rm e}^{-A} \approx \tfrac{a_1}{A'}$. We can then express all terms of ${\rm e}^A$, ${\rm e}^{-A}$ as $A'$ in our equations for $A''$ and $\phi''$. {\textcolor{black}  With this in hand, we are now able to} characterize the diverging behaviour of our equations in terms of just $A'$.

{\textcolor{black}
Returning to our expression for ${\rm e}^{B(r)}$, we substitute $E(r)$, $V(r)$ from \eqref{eqn:EF}, \eqref{eqn:EP} into \eqref{eqn:qd-eB} so that the equation's terms for $E$ and $V$ are now given by expressions of $Q$. We also express all instances of ${\rm e}^{\pm A}$ in terms of $A'$ instead using \eqref{eqn:NH-A1}. After rearranging this yields a new formulation for ${\rm e}^{B}$, which we use to do a series expansion of ${\rm e}^{B}$ about $A'$, to 3rd order. This yields,
}
\begin{align}
	e^{B(r)} & \approx
	-\frac{N^4 \phi'^2 + 2\phi'^2 r^2 N^2 + r^2 \left(\phi'^2 r^2 - 2 \right) \left( \beta \phi + \tfrac18 \right)}
	{2\left( \left( N^2 + r^2 \right) \left(\beta \phi + \tfrac18\right) - 4 Q^2 \right)}
	 \nonumber \\
	& \frac{
		8 \left(\beta \phi + \tfrac18 \right) \left(N^2 + r^2\right)^3
		\left(8 a_1 \alpha \phi' + 2 a_1 r \right)
	}
	{16 \left(N^2 + r^2 \right)^2 \left( \left(N^2 + r^2\right) \left(\beta \phi + \tfrac18 \right) - 4 Q^2 \right)}
	\left(A'\right)
	+ {\mathcal{O}}\left(\frac{1}{A'} \right).
	\label{eqn:hoz-B}
\end{align}
{\textcolor{black}
Higher order expansions only contribute terms proportional to ${\mathcal{O}}\left(\frac{1}{A'} \right)$ where $\frac{1}{A'} \to 0$ as $A' \to \infty$, and therefore they do not contribute meaningfully near the horizon as $r \to r_H$. 

Examining the terms proportional to $A'$ from \eqref{eqn:hoz-B}, we can identify some preliminary bounding behaviour. First, we require the numerator of this term to be finite in the near horizon limit so that solutions exist. Enforcing this requirement we have $r_H \leq - 4 \alpha \phi'$, which we interpret as an upper bound on $r_H$, as $\phi'$ must be negative for our scalar field to decay. Alternatively, this gives one potential approximate solution for $\phi'$ near the horizon, $\phi'_H = - \frac{r}{ 4 \alpha}$, in agreement with Einstein Gauss-Bonnet scalar solutions \cite{Bakopoulos}, and confirming $\phi'$ must be negative for any positive $\alpha$ on a real-valued domain. While this does not provide definite values, it offers some useful insights into the bounding behaviour of $\phi'$ and $r_H$.

Next we carry out this same procedure for our overall system of equations \eqref{eq:DE-A2}, \eqref{eq:DE-P2} with the previously included substitutions for ${\rm e}^{B}$ and $B'$. Performing a series expansion about $A'$ to 3rd order for these equations results in expressions of the form,
}
\begin{align}
	\begin{split}
	\phi'' &= f(r, \phi, \phi', A; \alpha, \beta) (A') + \mathcal{O}(1),
	\label{eqn:hoz-phi}
	\end{split} \\
	\begin{split}
	A'' &= g(r, \phi, \phi', A; \alpha, \beta) (A')^2 + \mathcal{O}(A'),
	\label{eqn:hoz-A}
	\end{split}
\end{align}
\noindent where $f$ and $g$ {\textcolor{black} represent expressions that are each composed of over a hundred-thousand terms, and are therefore} too large to reproduce in full here. {\textcolor{black} Doing a series expansion to 4th order only yield terms proportional to ${\cal O}(1)$ in the case of \eqref{eqn:hoz-phi} and no additional terms for \eqref{eqn:hoz-A}. As we are only interested in the divergent behaviour of these functions, these additional terms do not meaningfully contribute to our approximations near the horizon. While these expressions are too large to use analytically, we were able to evaluate them numerically for all parameters explored in the following sections. Estimations of initial conditions did not vary outside numerical precision between 3rd and 4th order.}

For $A'$ and $\phi'$ to have solutions near the horizon, $A''$ and $\phi''$ need to be finite as $r \rightarrow r_H$. Imposing this condition on $f$, we find an expression for $\phi'$ such that $f=0$, leaving $\phi''$ finite as $\phi$ diverges. We find three valid solutions for this $\phi'_H$, the simplest is $\phi'_H=-\tfrac{r}{4 \alpha}$, {\textcolor{black} consistent with our results from \eqref{eqn:hoz-B},} but this does not sufficiently describe the behaviour of $\phi'$ in our case. Looking at the remaining two solutions we find that they have the form of, 
\begin{equation}
	\phi'_H = \frac{\zeta \pm \eta \sqrt{\kappa}}{\xi}.
\end{equation}
\noindent Given the size and complexity of $f$, the solutions for $\phi'_H$ are too large to be reproduced in this paper. However, we find that all its components, $\zeta$, $\eta$, $\kappa$, and $\xi$ reduce down to the results of \cite{Hunter} when $N=0$, and we also recover the results of the Einstein-scalar-Gauss-Bonnet case \cite{Sotiriou} when we also set $Q, \beta = 0$.
There are no restrictions on $\phi$ and its initial value is a free parameter that will be set as required by our numerical approach. In our case, $\phi_H'$ retains terms of the expansion coefficient $a_1$. The choice of $a_1$ has no impact on $\phi_H'$ within numerical precision, so long as $a_1$ is not near zero.

Finally, we note that for all parameter values examined $g \approx \tfrac12$ for any $|a_1| >> 0$, which guarantees $A''$ remains finite as $A$ diverges. This allows us to simply treat the function $g$ as a constant for all examined parameters. Integration of \eqref{eqn:hoz-A} is then straight forward, giving us the approximations $A' \approx \tfrac{1}{2 (r - r_H)}$ and $A \approx \ln[2 (r - r_H)]$. We assume any constants from integration are negligible when $r$ is sufficiently close to $r_H$.

\subsection{Asymptotic Limit}

Next we examine the asymptotic limits as $r \rightarrow \infty$. As with the near horizon limit, we have known expectations of how our functions should behave in the limiting case. We would expect that the metric solution must converge to the Schwarzschild and Reissner–Nordström metric solutions in this limit. That requires that ${\rm e}^{A}, {\rm e}^{B} \rightarrow 1$ as $r \rightarrow \infty$. Additionally, the scalar field $\phi$ should converge to a finite value in this limit, and by extension $\phi' \rightarrow 0$. The electric field $E$ should also converge to zero to satisfy the Maxwell equations, like any other static radial electric field examined in the limit of $r \rightarrow \infty$.

We now take a perturbation approach to explore this limiting case by constructing power series expansions for each function around $\tfrac{1}{r}$, giving us the following,
\begin{subequations}
\begin{align}
\begin{split}
	{\rm e}^{A(r)} &= 1 + \sum_{n=1}^{\infty} \frac{c^{A}_{n}}{r^{n}},
\end{split}\\
\begin{split}
	{\rm e}^{B(r)} &= 1 + \sum_{n=1}^{\infty} \frac{c^{B}_{n}}{r^{n}},
\end{split}\\
\begin{split}
	\phi(r) &= \sum_{n=1}^{\infty} \frac{c^{\phi}_{n}}{r^{n}},
\end{split}\\
\begin{split}
	{\rm E}(r) &= \sum_{n=2}^{\infty} \frac{c^{E}_{n}}{r^{n}}.
\end{split}
\end{align}
\end{subequations}
\noindent Substituting these into our differential equations, \eqref{eqn:DE-Gtt}, \eqref{eqn:DE-Grr}, \eqref{eqn:DE-Gth}, \eqref{eqn:DE-Scalar} we can solve for the series coefficients $c^A_n$, $c^B_n$, $c^{\phi}_n$, $c^E_n$, order-by-order. 

We set the 1st order constants for ${\rm e}^A$, ${\rm e}^B$ such that $c^{A}_{1} = -2M$, $c^{B}_{2} = 2M$ to correspond with Schwarzschild metric and $c^{\phi}_{1}=P$ to indicate the scalar charge of the black hole. Additionally, the 1st term for $E$ is set as $c^{E}_{2} = Q$, where $Q$ is the electric charge of the black hole. All together, this gives the following perturbation approximations,
\begin{subequations}
\begin{align}
\begin{split}
	{\rm e}^{A(r)} &= {\scriptstyle 1} - \tfrac{2\, M}{r} + \tfrac{{Q^2 - 2\,N^2}}{4 \,r^2}
	+ \tfrac {12\,M{N}^{2} + M P^2 - 4 \, P Q^2 \beta}{6 r^3} + {\cal O}(\tfrac{1}{r^4}),
\end{split}\\
\begin{split}
	{\rm e}^{B(r)} &= {\scriptstyle 1} + \tfrac{2\, M}{r} + \tfrac{8\,M^2 + 4\,N^2 - P^2 - Q^2}{2\, r^2}
	+ \tfrac{16\,M^3 + 12\,MN^2 + 4\,PQ^2 \beta  - 5\,MP^2 - 2\,MQ^2}{2\, r^3} + {\cal O}(\tfrac{1}{r^4}),
\end{split}\\
\begin{split}
	\phi(r) &= \tfrac{P}{r} + \tfrac{MP - Q^2\beta }{r^2} 
	+ \tfrac{64\,PQ^2 \beta^2 - 16\,MQ^2\beta + 16\,PM^2 + 4\,N^2P - P^3 - PQ^2}{12\, r^3} + {\cal O}(\tfrac{1}{r^4}),
\end{split}\\
\begin{split}
	{\rm E}(r) &= \tfrac {Q}{r^2} - \tfrac {8 P \beta \,Q}{r^3} + \tfrac {256\,P^2Q\beta^2 + 32\,Q^3\beta^2 - 32\,MPQ\beta - 2\, M^2Q - 24\,N^2 Q + P^2Q + Q^3 }{4\, r^4} + {\cal O}(\tfrac{1}{r^5}).
\end{split}
\end{align}
\label{eqn:pertBCs}
\end{subequations}
Here we report the first 3 orders for each term. We find that in 1st order we recover Schwarzchild for ${\rm e}^A$ and ${\rm e}^B$. The scalar field converges w.r.t. $\tfrac{P}{r}$ and the electric field to the familiar $\tfrac{Q}{r^2}$ term, as required. One additional order was also calculated for each function, however for our numerical approach the additional term did not contribute to the boundary conditions within numerical precision, and therefore was not needed when calculating numerical values for our boundary conditions.

\subsection{Numerical Approach}

Our numerical approach utilizes the Julia \texttt{DifferentialEquations} \cite{Julia-DE} and \texttt{Optim} \cite{Julia-Opt} packages. For this problem, we employ a shooting method to treat our boundary value problem (BVP) as an initial value problem (IVP). As we are ``shooting'' to solve two functions whose initial conditions are given by approximations of their divergences near the horizon, standard algorithms often struggle to successfully iterate the initial conditions for future shooting attempts. To address this, we used a  modification of the shooting BVP solver provided in \texttt{DifferentialEquations}, where we supplement the non-linear solver for the optimization algorithms from the \texttt{Optim} package. This allowed us to tune the shooting approach to be suitable for gravitational problems of this kind, with initial conditions approximating diverging functions. 

For the numerical integrator, we choose a 2nd order A-stable symmetric ESDIRK implicit method, \texttt{Trapezoid()} \cite{Julia-DE}, which offers a good balance between computational cost and accuracy. This Trapezoid integrator is based on the spatial step of the Crank-Nicholson algorithm used for PDEs. Other implicit methods may offer additional advantages but care needs to be taken as many implementations will employ ancillary methods ill-suited to this type of problem. Different approaches to extrapolating implicit values on the first step or algorithms that use smoothing techniques can interact with the divergences in unexpected ways. Notably, \texttt{Trapezoid()} does not use numerical dampening, which can have unpredictable results in our case. While explicit methods were tried, those integrators were too sensitive to the initial conditions to be reliable for all parameter sets utilized for our solutions in the following section.

All numerical calculations start at $r=1.01$ determined by sampling various $r$ values. We assume the horizon radius to be $r_H=1.0$. The {\em true} horizon will be at some $r_{TH} \leq r_{H}$ and due to the complexity of our equations calculating a meaningful bounding conditions on $r_{TH}$ is not tractable. The numerical calculations can be started from $r_{H} < r < 1.01$ but this offers no meaningful benefits and merely increases computation time with no useful change to result accuracy. We also take the numerical integrations to $r=10^{8}$, which sees all functions converge to the asymptotic boundary condition within tolerance. All numerical integrations were done with a relative tolerance of $10^{-14}$ and an absolute tolerance of $10^{-8}$, which offers the best balance between accuracy and time cost.

We can evaluate the error of our solutions by substituting them back into the original equations of motion. The scalar field equation \eqref{eqn:DE-Scalar} notably acts as a good check of our solutions as it was not directly used formulating the equations being solved. The only equation unavailable for us to use is $G_{rr}$, as we rearranged it to solve for and substitute away terms of $B$ and $B'$. Thus our solutions depend on implicit integrations of $B'$. This limits our ability to produce $B$ explicitly using \eqref{eqn:de-qd-eb} or to use $G_{rr}$ for error calculations.

However, we may still utilize $G_{tt}$ \eqref{eqn:DE-Gtt}, $G_{\theta\theta}$ \eqref{eqn:DE-Gth}, the scalar field equation \eqref{eqn:DE-Scalar} and the 2 independent equations from \eqref{eqn:FE-MW} without issue. The only equation to not be identically zero within numerical precision for any result was $G_{tt}$, and only very near the horizon. Across all reported solutions error is at most on the order of $\epsilon = 10^{\textcolor{black}{-5}}$ or ${\cal O}(0.001\%)$ and only for $1.01 \leq r < 1.1$. Error is always zero within numerical precision past this region.

We employ an Eddington–Finkelstein like coordinate change such that,
\begin{equation}
	x = 1 - \frac{r_H}{r},
\end{equation}
which maps every $r$ to an $x$ on the closed interval $[0, 1]$. Numerically this gives an open range of $x=[0.099, 0.999]$. Most of the numerical range covers a region close to the horizon, e.g. $r=20 \mapsto x=0.95$. This naturally biases any solver to be more granular in the region where accuracy is most important for our problem.

Finally, every calculation is defined by the parameters $P$, $M$ which determines the boundary conditions calculated with \eqref{eqn:pertBCs}. We initially assume $P, M = 1.0$ for all calculations. Both boundaries and the system of equations depend on $N$, $Q$, $\alpha$, $\beta$. We establish $N>0$ from our metric and we assume $Q, \alpha, \beta > 0$. The following results explore variations of our parameters within these constraints. 

\section{Results}\label{sec:results}

\subsection{Varying Electric Charge Q} \label{sec:Results-VaryQ}

We begin by examining our results when electric charge $Q$ is varied. We consider values of $Q$ that produce real-valued solutions only. Any integration attempts that became complex-valued anywhere on the domain were discarded algorithmically. To ensure real-valued solutions, we found that the relationship between $Q$ and the Maxwell coupling constant $\beta$ had to be taken into account. To identify suitable values of $Q$ we examine the behaviour of $\phi_H'$ with respect to the electric charge for various values of $\beta$. We chose $\phi_H'$ as $f$ in \eqref{eqn:hoz-phi} is the only component of our system of equations that has a strong $Q$ dependence near the horizon. We find that $\phi_H'$ demonstrates a phase change as $Q$ is increased to sufficiently large values for a given choice of $\beta$. Other parameters such as $\alpha$ and $N$ contribute to the behaviour of $\phi_H'$ but only do so to a far lesser degree and they do not interact with other parameters.
\begin{figure}[h]
	\captionsetup[subfigure]{justification=centering}
	\begin{subfigure}{.42\textwidth}
		\includegraphics[width=\linewidth]{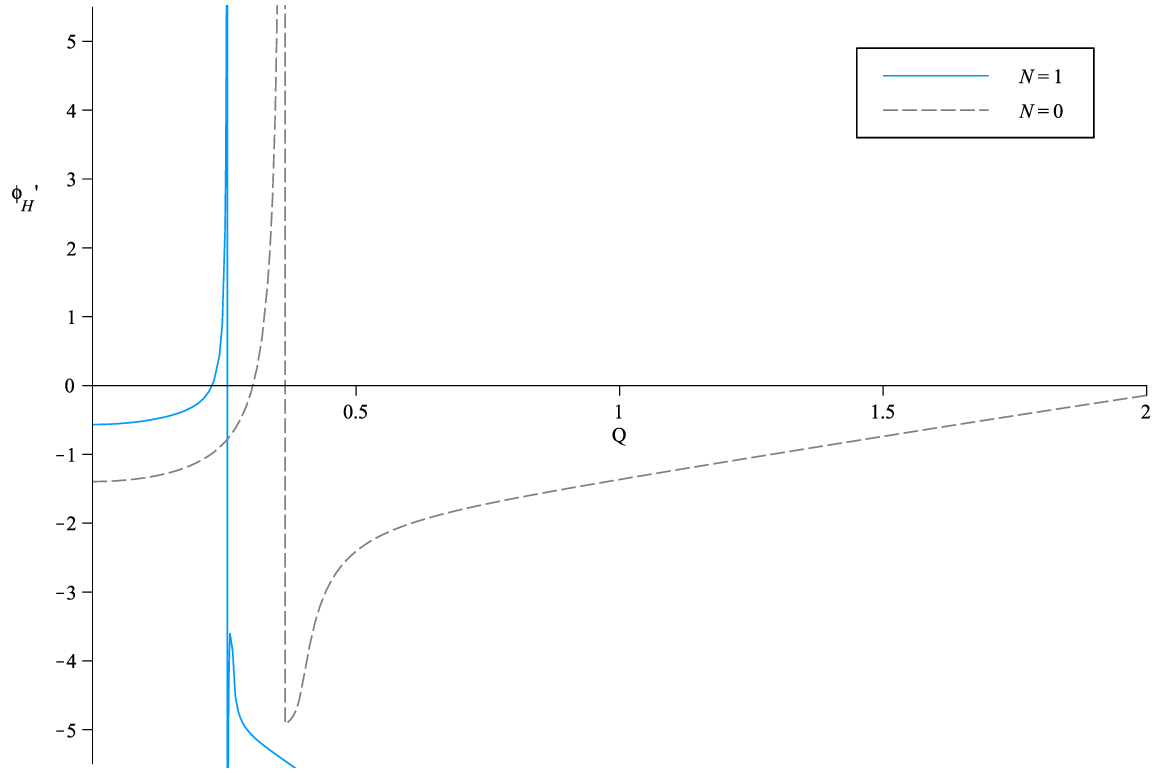}
		\subcaption{$\beta=0.05$}
		\label{fig:P1H_QBeta005}
	\end{subfigure}
	\begin{subfigure}{.42\textwidth}
		\includegraphics[width=\linewidth]{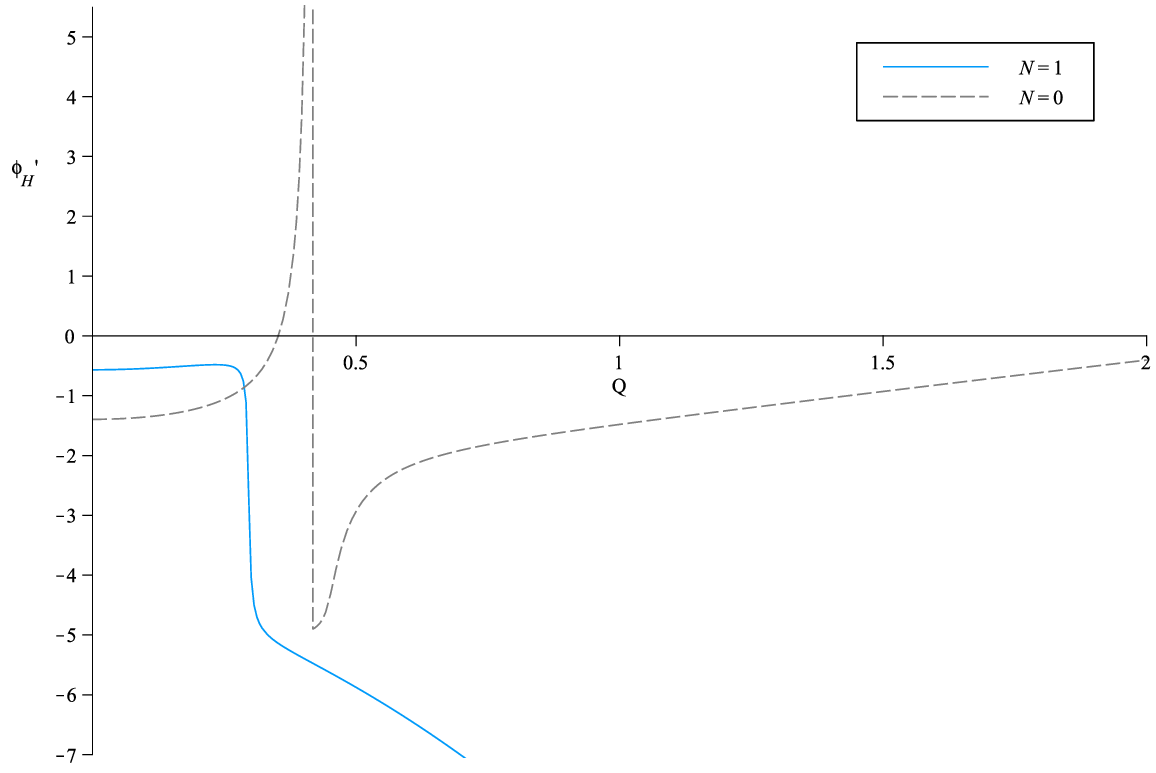}
		\subcaption{$\beta=0.1$}
		\label{fig:P1H_QBeta01}
	\end{subfigure}\\
	\begin{subfigure}{.42\textwidth}
		\includegraphics[width=\linewidth]{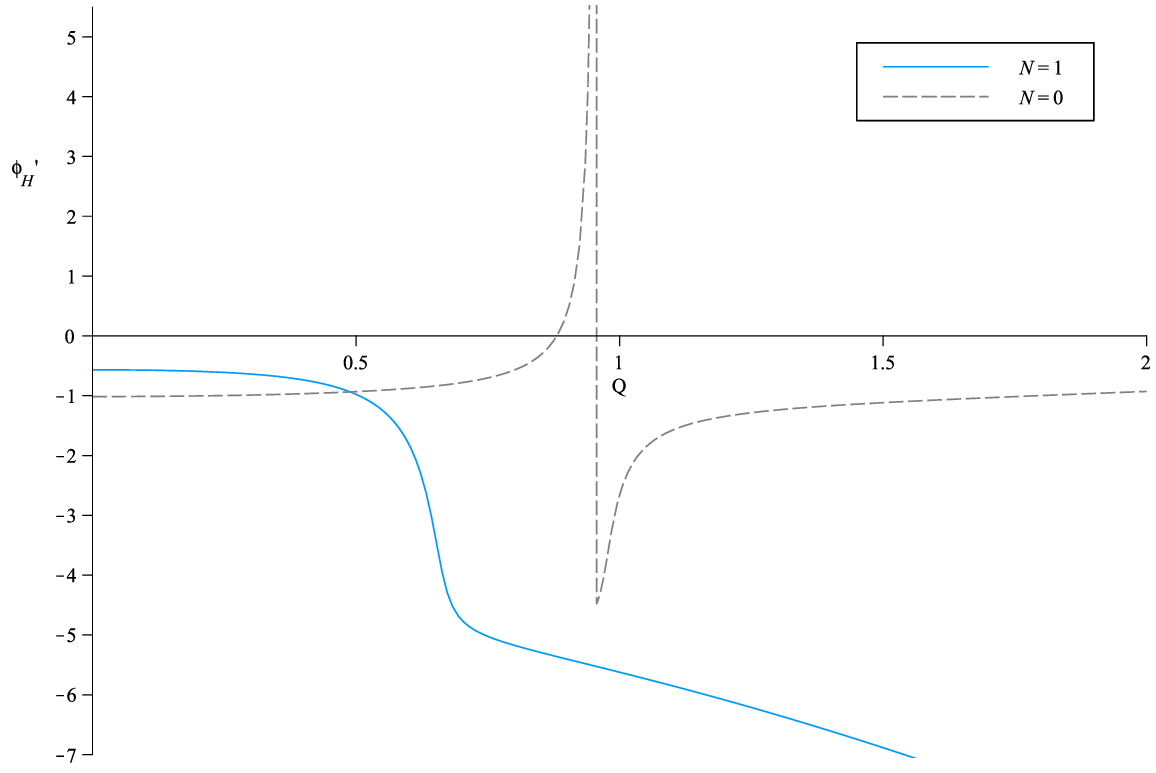}
		\subcaption{$\beta=1.0$}
		\label{fig:P1H_QBeta1}
	\end{subfigure}
	\begin{subfigure}{.42\textwidth}
		\includegraphics[width=\linewidth]{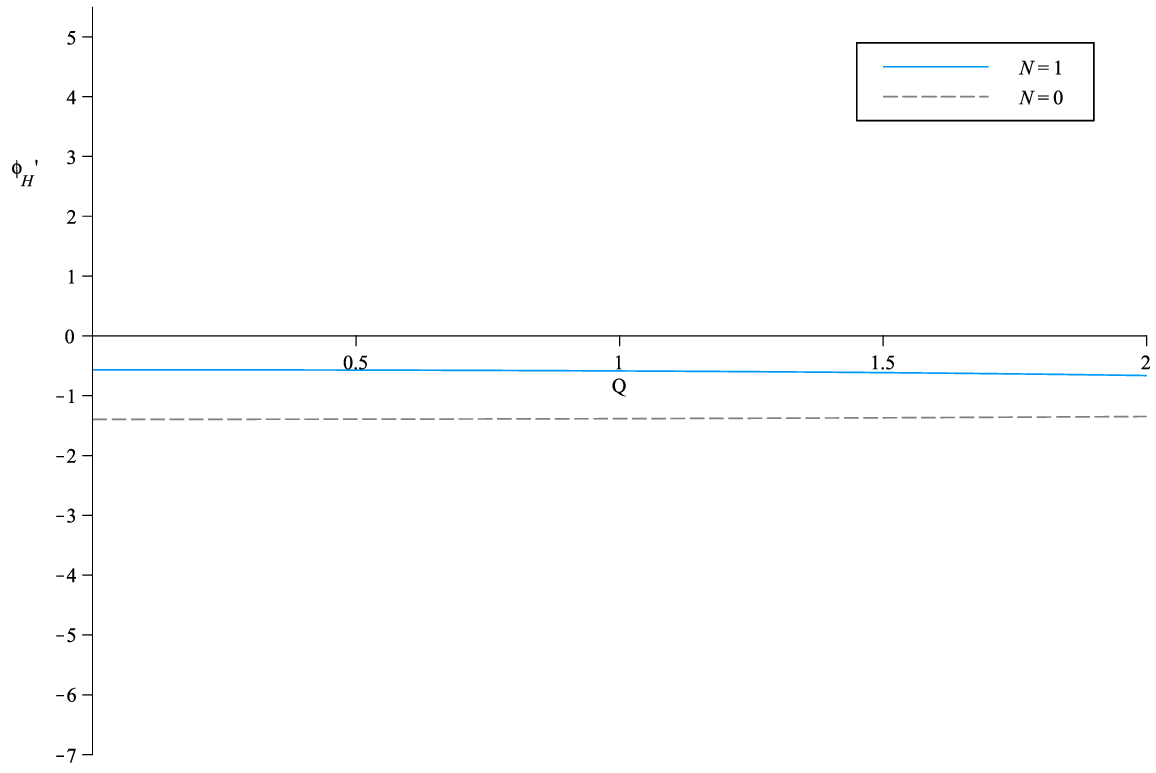}
		\subcaption{$\beta=50.0$}
		\label{fig:P1H_QBeta50}
	\end{subfigure}
	\caption{Behaviour of $\phi_{h}^{'}$ while varying $Q$ for multiple $\beta$ values, with $\alpha=0.05$, and $N=0$. Dotted lines (grey) represent the $N=0$ case and solid lines (blue) show the case with a NUT charge present.}
	\label{fig:P1H_QBeta}
\end{figure}

Looking at Figure \ref{fig:P1H_QBeta005}, we set $\beta=0.05$ and observe a phase change occurs around $Q\approx0.25$ with the NUT charge present, and at $Q\approx0.36$ with $N=0$. We find all solutions starting with a $Q$ value to the RHS of this phase change are numerically unstable or complex-valued. If we increase the coupling constant $\beta$, as we did in Figures \ref{fig:P1H_QBeta01}, \ref{fig:P1H_QBeta1}, \ref{fig:P1H_QBeta50} we can note that the larger coupling constant causes the phase change to shift towards higher values of $Q$. Therefore, with a sufficiently large choice of $\beta$ the phase change moves out of the region of interest for the electric charge.

\begin{figure}[h]
	\captionsetup[subfigure]{justification=centering}
	\begin{subfigure}{.42\textwidth}
		\centering
		\includegraphics[width=\linewidth]{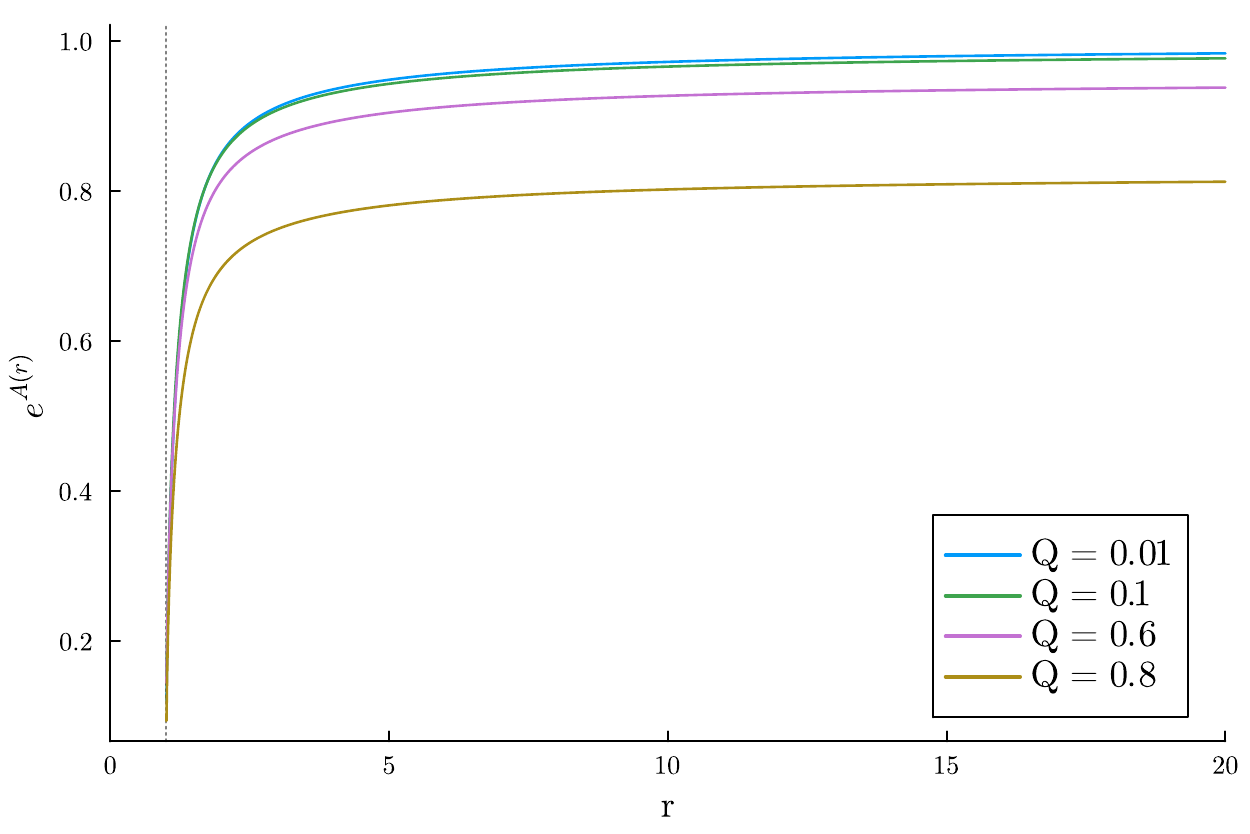}
		\vspace{-1\baselineskip}
		\caption{Results for $\rm{e}^{A(r)}$ while varying $Q$ for $\alpha=0.03$, $\beta=30.0$, and $N=1.0$.}
		\label{fig:VaryQa}
	\end{subfigure}
	\begin{subfigure}{.42\textwidth}
		\centering
		\includegraphics[width=\linewidth]{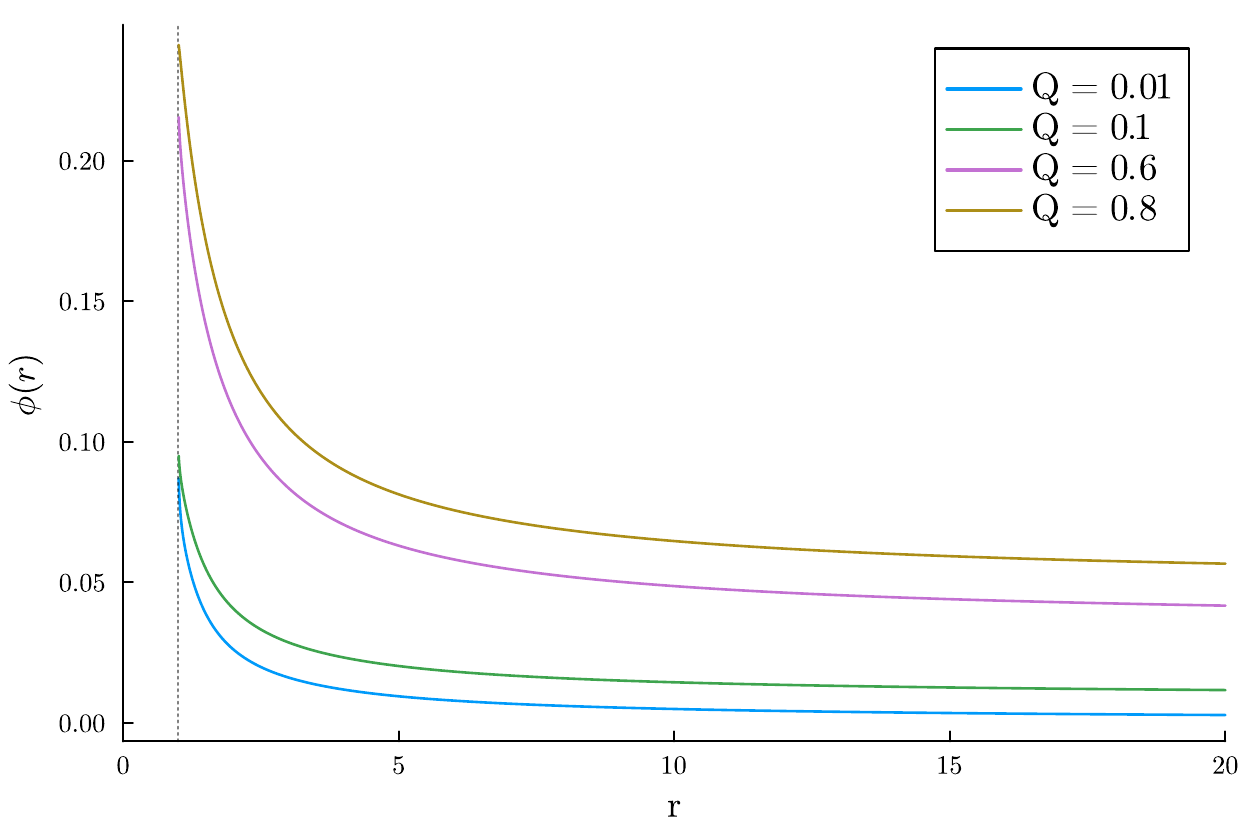}
		\vspace{-1\baselineskip}
		\caption{Results for $\phi(r)$ while varying $Q$ for $\alpha=0.03$, $\beta=30.0$, and $N=1.0$.}
		\label{fig:VaryQb}
	\end{subfigure}
	\caption{Solutions for $\alpha=0.03$, $\beta=30.0$, and $N=1.0$ while varying $Q$.}
	\label{fig:VaryQ}
\end{figure}

With suitable values for $\beta$ identified we may now examine our results for varying $Q$. We choose to explore solutions in the range $Q=[0.01, 0.8]$. Starting with Figure \ref{fig:VaryQa}, we have solutions for ${\rm e}^A$, the time solution for metric. We find ${\rm e}^A$ increases in magnitude with increasing values of $Q$, and widens the difference from ${\rm e}^A \rightarrow 1$ for a given large but finite value of $r$, as expected from \eqref{eqn:pertBCs}. In a similar fashion, in Figure \ref{fig:VaryQb} the magnitude of the scalar field $\phi$ is found to increase with increasing values in $Q$ with a wider difference from the zero boundary out towards infinity. Therefore, as electric charge increases we see in both cases that the magnitude of our solutions increases, and the rate of convergence to the far boundary value is slowed.

\begin{figure}[h]
	\captionsetup[subfigure]{justification=centering}
		\begin{subfigure}{.425\textwidth}
		\centering
		\includegraphics[width=\linewidth]{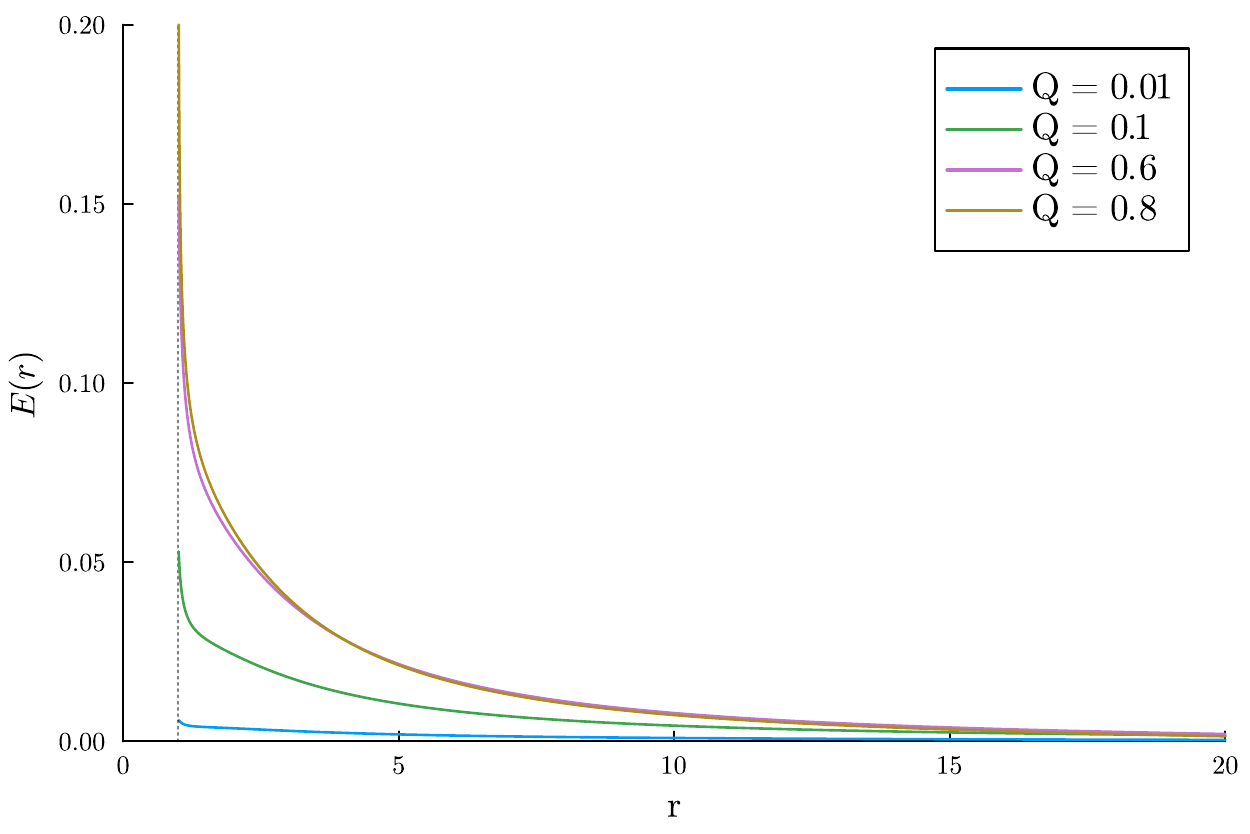}
	\end{subfigure}
	\begin{subfigure}{.425\textwidth}
		\centering
		\includegraphics[width=\linewidth]{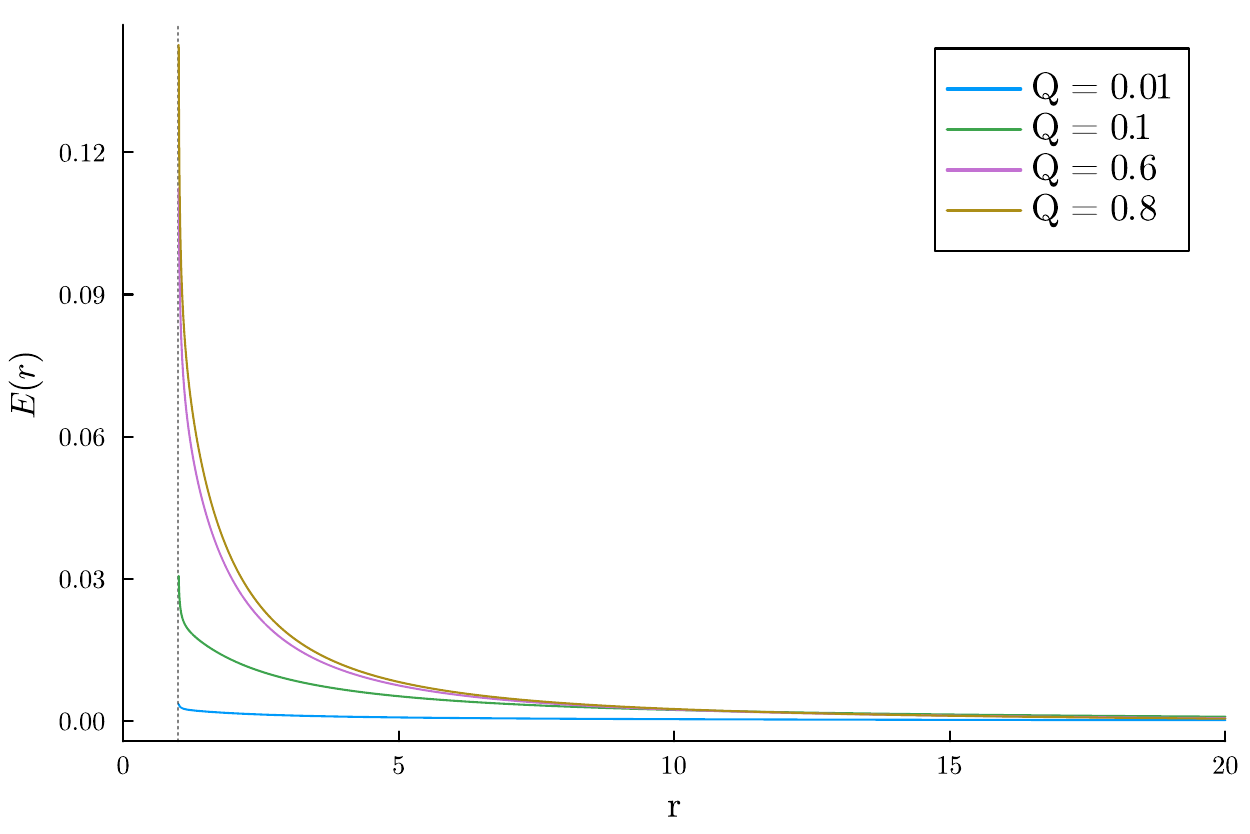}
	\end{subfigure}
	\begin{subfigure}{.425\textwidth}
		\centering
		\includegraphics[width=\linewidth]{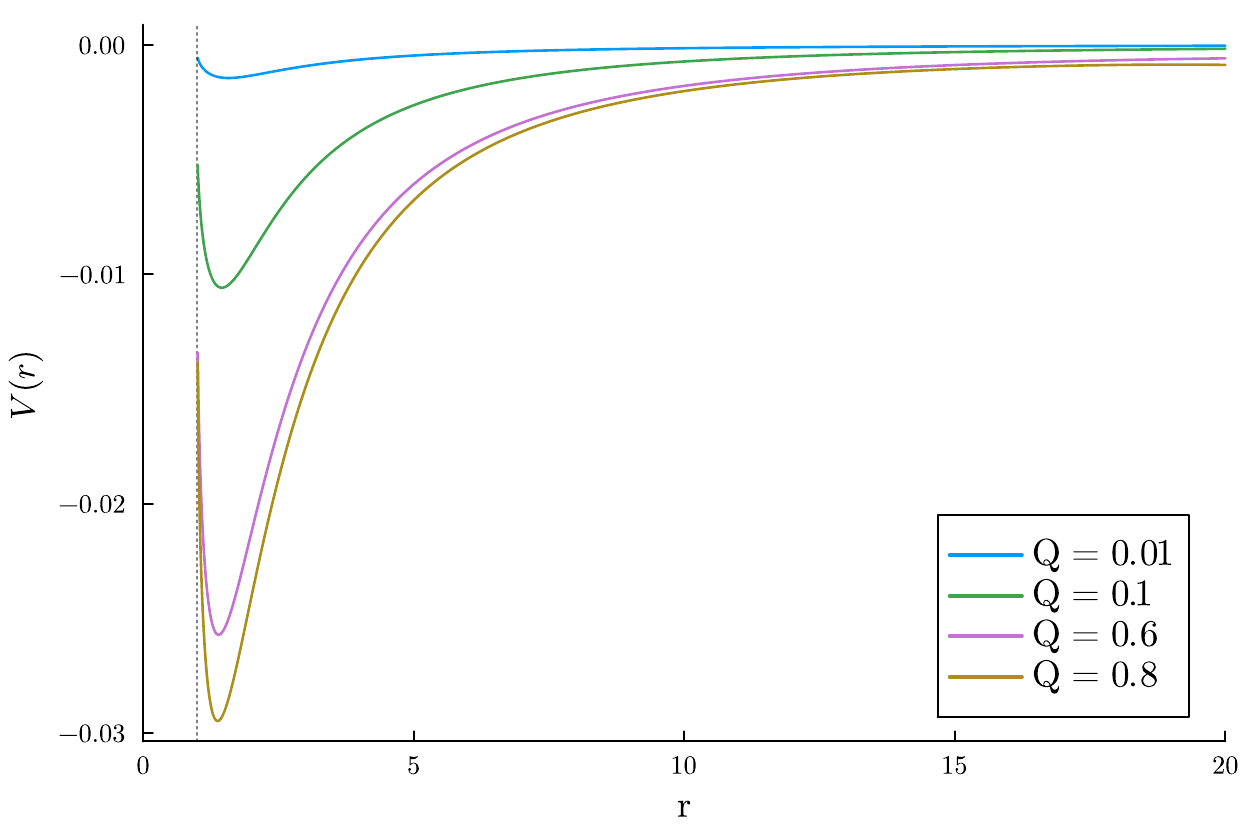}
		\caption*{$N=1$}
	\end{subfigure}
	\begin{subfigure}{.425\textwidth}
		\centering
		\includegraphics[width=\linewidth]{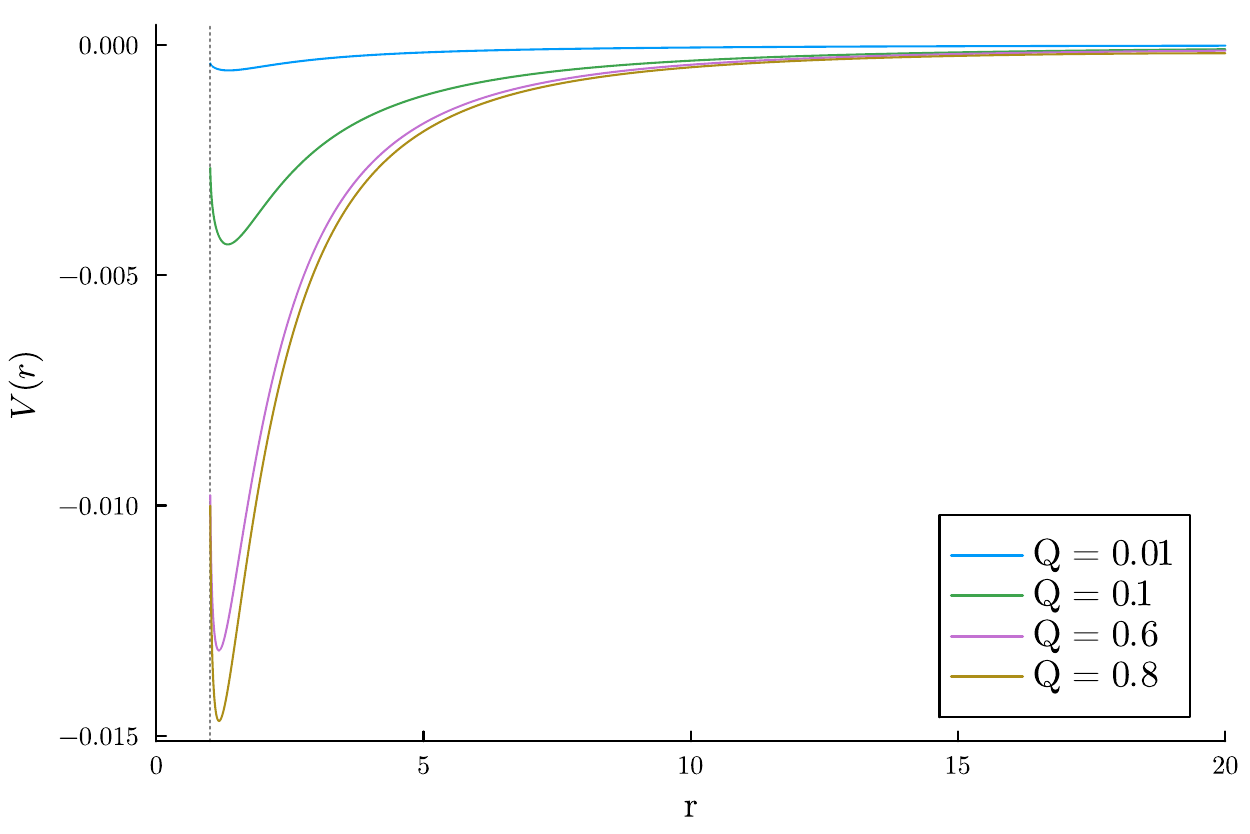}
		\caption*{$N=0$}
	\end{subfigure}
	\caption{Results for electric field $E(r)$ and electric potential function $V(r)$ while varying $Q$ for $\alpha=0.03$, and $\beta=30.0$.}
	\label{fig:VaryQ-EF}
\end{figure}

In Figure \ref{fig:VaryQ-EF}, we find that the electric field $E$ increases with increasing charge, as required. Comparing these results to the $N=0$ case, $E$ is stronger in the presence of $N$. The electric potential function $V$, increases in magnitude for increasing $Q$. In the presence of $N$, we see approximately a doubling in the strength compared to the $N=0$ case. As $N$ is an effective magnetic charge and contributes to the potential, this behaviour agrees with our expectations.

\subsection{Varying $\alpha$}

Next we examine the relationship that $\alpha$ has on our solutions. We look at values of $0.01 \leq \alpha \leq 0.08$. While a choice of $\alpha < 0.01$ can be computed and does produce real-valued initial conditions, it does not produce any interesting test cases as we are suppressing meaningful contributions from the Gauss-Bonnet term. As we did for varying Q in Section \ref{sec:Results-VaryQ}, we examine the behaviour of $\phi_H'$ to identify appropriate values of $\alpha$. We found that $\phi_H'$ becomes complex-valued for $\alpha \ge 0.08$, therefore we only consider values below this threshold.

\begin{figure}[h]
	\captionsetup[subfigure]{justification=centering}
	\begin{subfigure}{.45\textwidth}
		\centering
		\includegraphics[width=\linewidth]{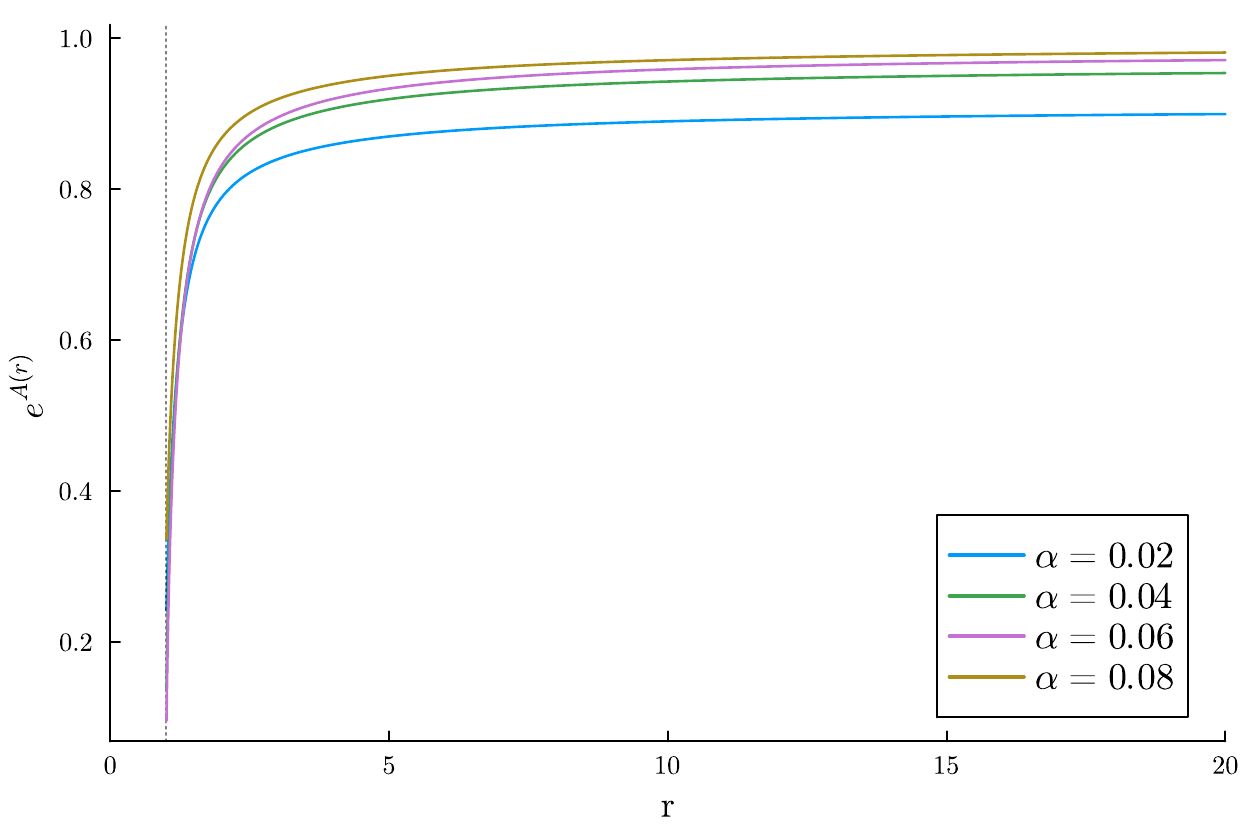}
		\caption{}
		\label{fig:VaryAa}
	\end{subfigure}
	\begin{subfigure}{.45\textwidth}
		\centering
		\includegraphics[width=\linewidth]{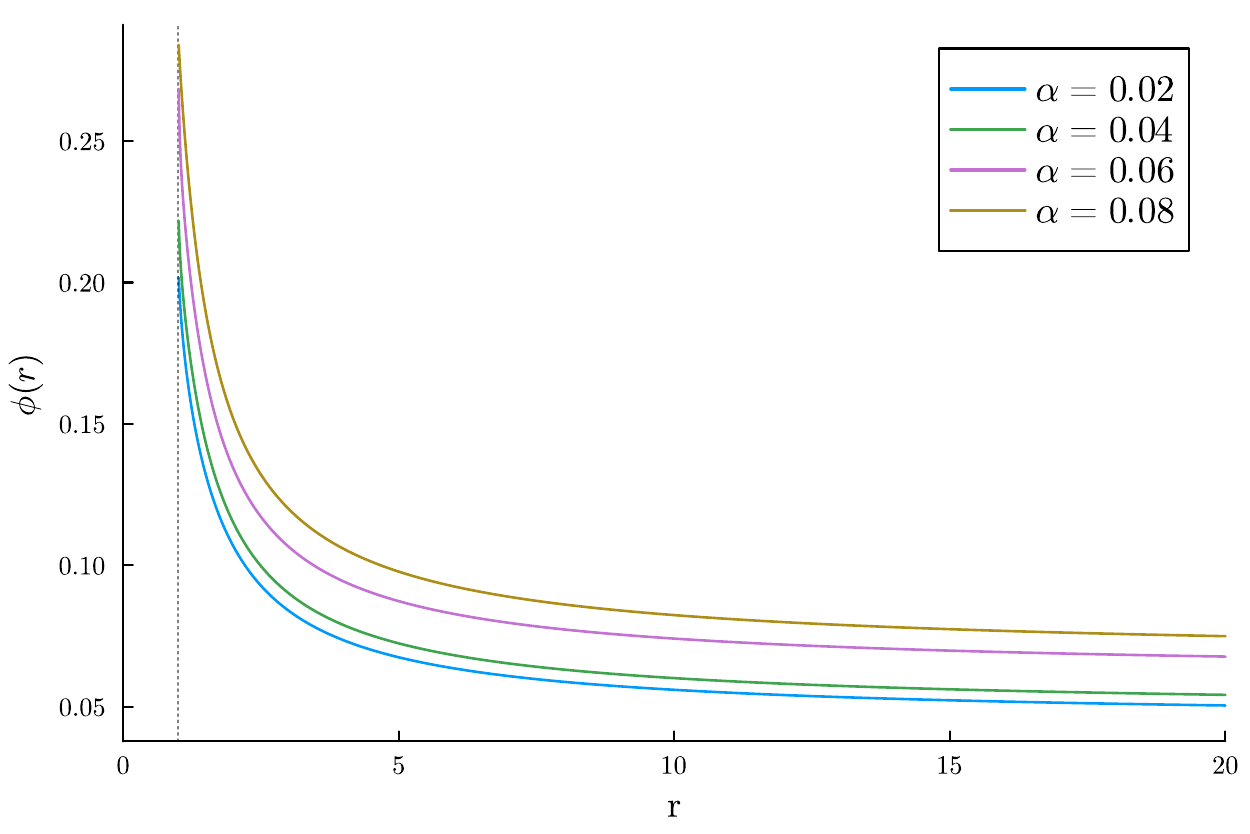}
		\caption{}
		\label{fig:VaryAb}
	\end{subfigure}
	\caption{Results for $e^{A(r)}$ and $\phi(r)$ while varying $\alpha$ for $Q=0.5$, $\beta=30.0$, and $N=1.0$.}
	\label{fig:VaryA}
\end{figure}

In figure \ref{fig:VaryAa}, we see that $\alpha$ and ${\rm e}^A$ have an inverse relationship, with the magnitude of ${\rm e}^A$ decreasing as $\alpha$ is increased. This aligns as to why we see an upper limit for $\alpha > 0.8$ in our analysis, as increasing $\alpha$ is bounded by the asymptotic limit of ${\rm e}^A$. Conversely, in figure \ref{fig:VaryAb} we observe that the scalar field $\phi$ increases in magnitude with $\alpha$.

\begin{figure}[h]
	\captionsetup[subfigure]{justification=centering}
	\begin{subfigure}{.45\textwidth}
		\centering
		\includegraphics[width=\linewidth]{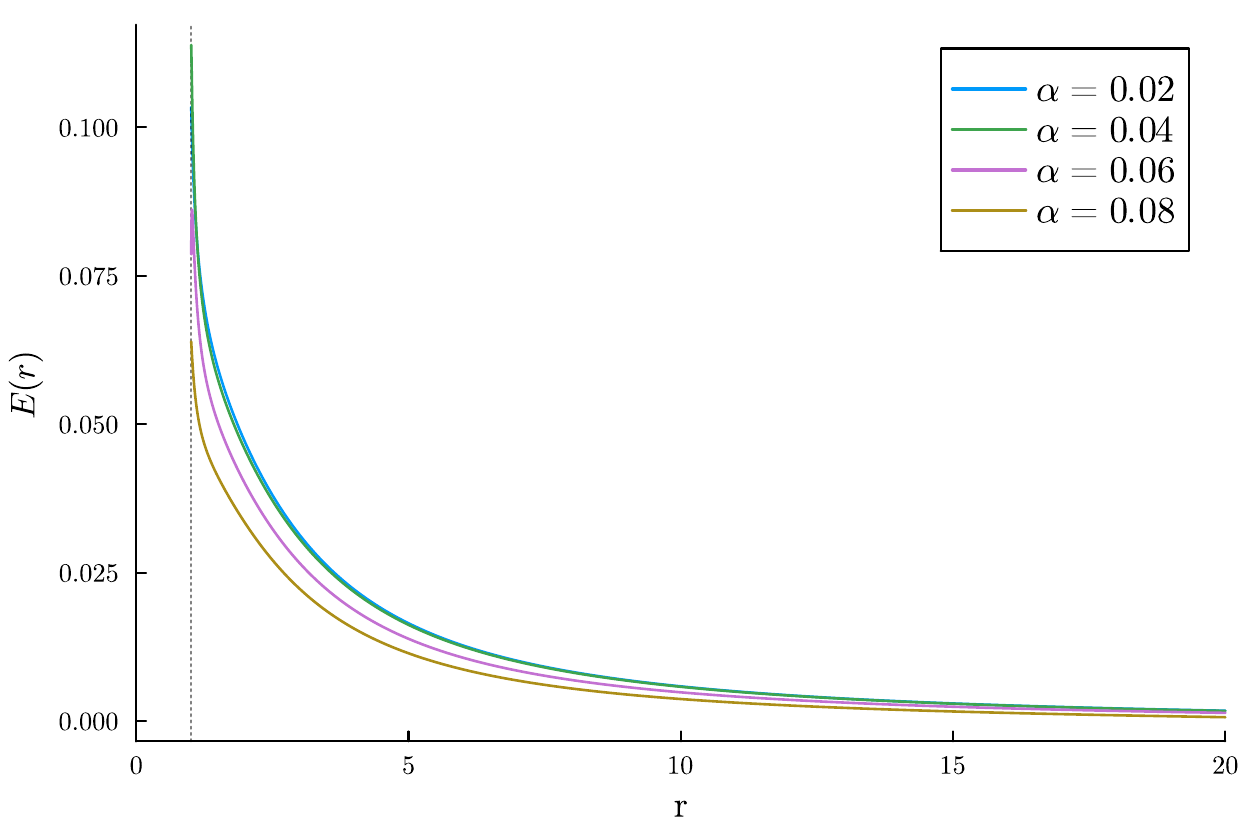}
		\caption{}
		\label{fig:VaryA-E}
	\end{subfigure}
	\begin{subfigure}{.45\textwidth}
		\centering
		\includegraphics[width=\linewidth]{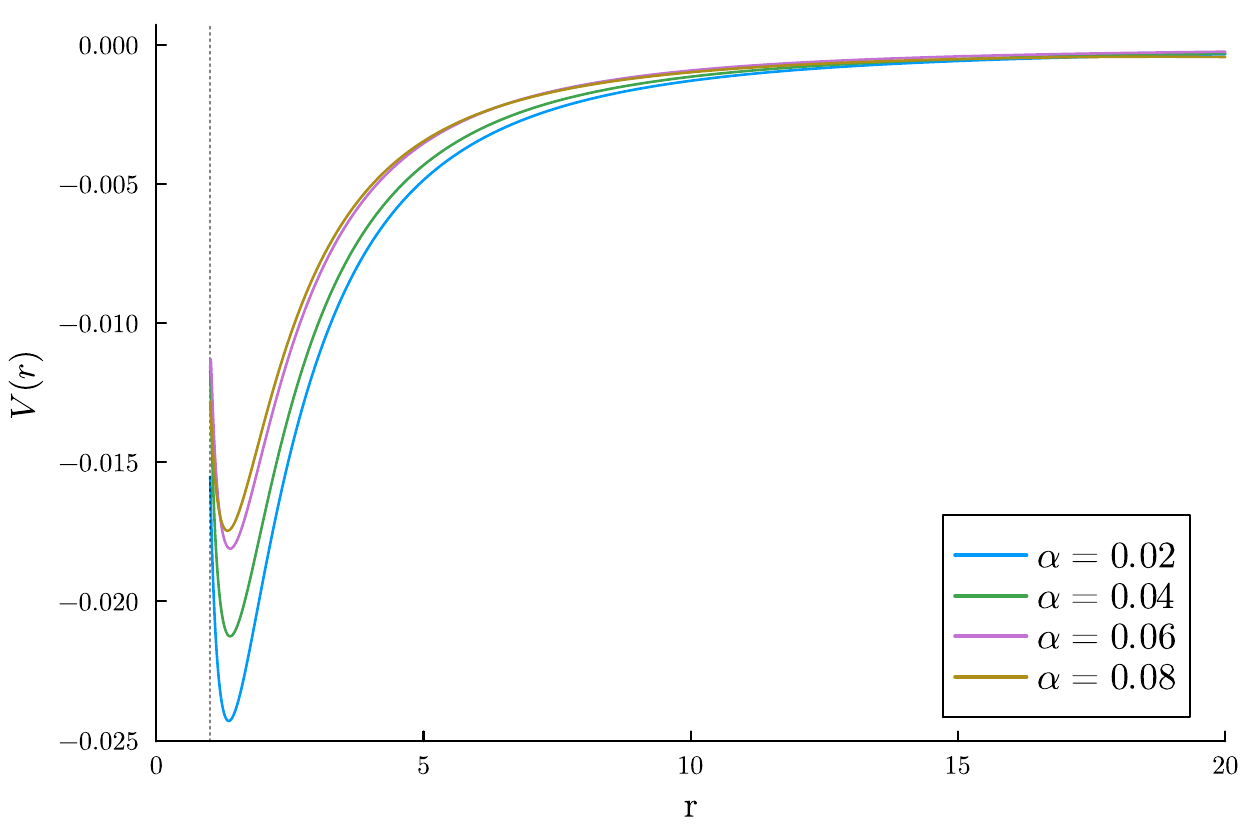}
		\caption{}
		\label{fig:VaryA-V}
	\end{subfigure}
	\caption{Results for $E(r)$ and $V(r)$ while varying $\alpha$ for $Q=0.5$, $\beta=30.0$, and $N=1.0$.}
	\label{fig:VaryA-EF}
\end{figure}

Finally, examining the electric field in Figure \ref{fig:VaryA-E} we find that $E$ is largely unaffected for smaller values of $\alpha$. Increasing values of $\alpha$ begin to increasingly suppress the electric field. We expect $E$ and $V$ to vary in magnitude proportional to each other, which observe in \ref{fig:VaryA-V}.

\subsection{Varying $\beta$}

Next we examine the behaviour of our solutions as we vary $\beta$ using values of $\alpha=0.03$, $Q=0.5$, and $N=1.0$. 
\begin{figure}[h]
	\captionsetup[subfigure]{justification=centering}
	\begin{subfigure}{.45\textwidth}
		\centering
		\includegraphics[width=\linewidth]{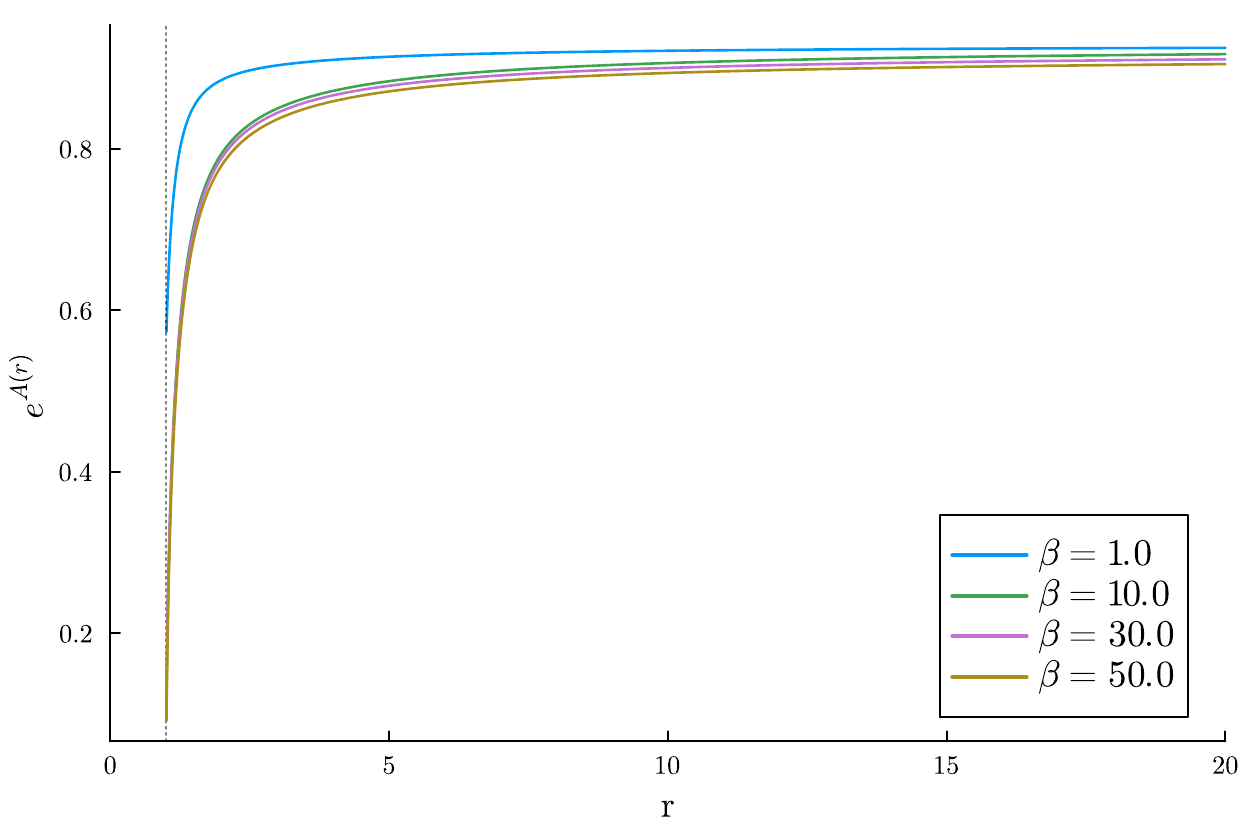}
		\caption{}
		\label{fig:VaryBe_a}
	\end{subfigure}
	\begin{subfigure}{.45\textwidth}
		\centering
		\includegraphics[width=\linewidth]{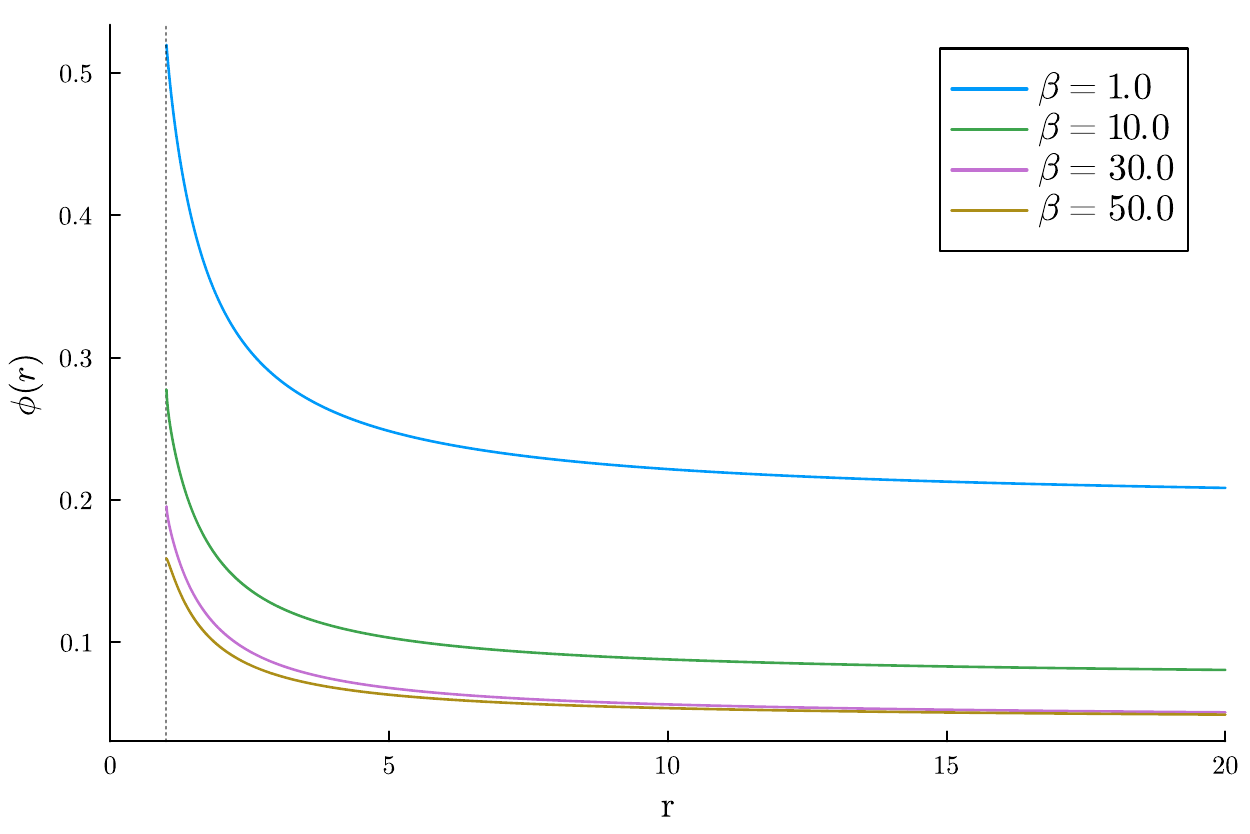}
		\caption{}
		\label{fig:VaryBe_b}
	\end{subfigure}
	\caption{Results for $e^{A(r)}$ and $\phi(r)$ while varying $\beta$ for $Q=0.5$, $\alpha=0.03$, and $N=1.0$.}
	\label{fig:VaryBe}
\end{figure}
The time component given in Figure \ref{fig:VaryBe_a} shows larger variance for smaller $\beta$ but is largely unaffected for larger values. We find that the scalar field in Figure \ref{fig:VaryBe_b} is strongly influenced by $\beta$, where larger $\beta$ suppresses $\phi$ demonstrating an inversely proportional relationship.
\begin{figure}[h]
	\captionsetup[subfigure]{justification=centering}
	\begin{subfigure}{.45\textwidth}
		\centering
		\includegraphics[width=\linewidth]{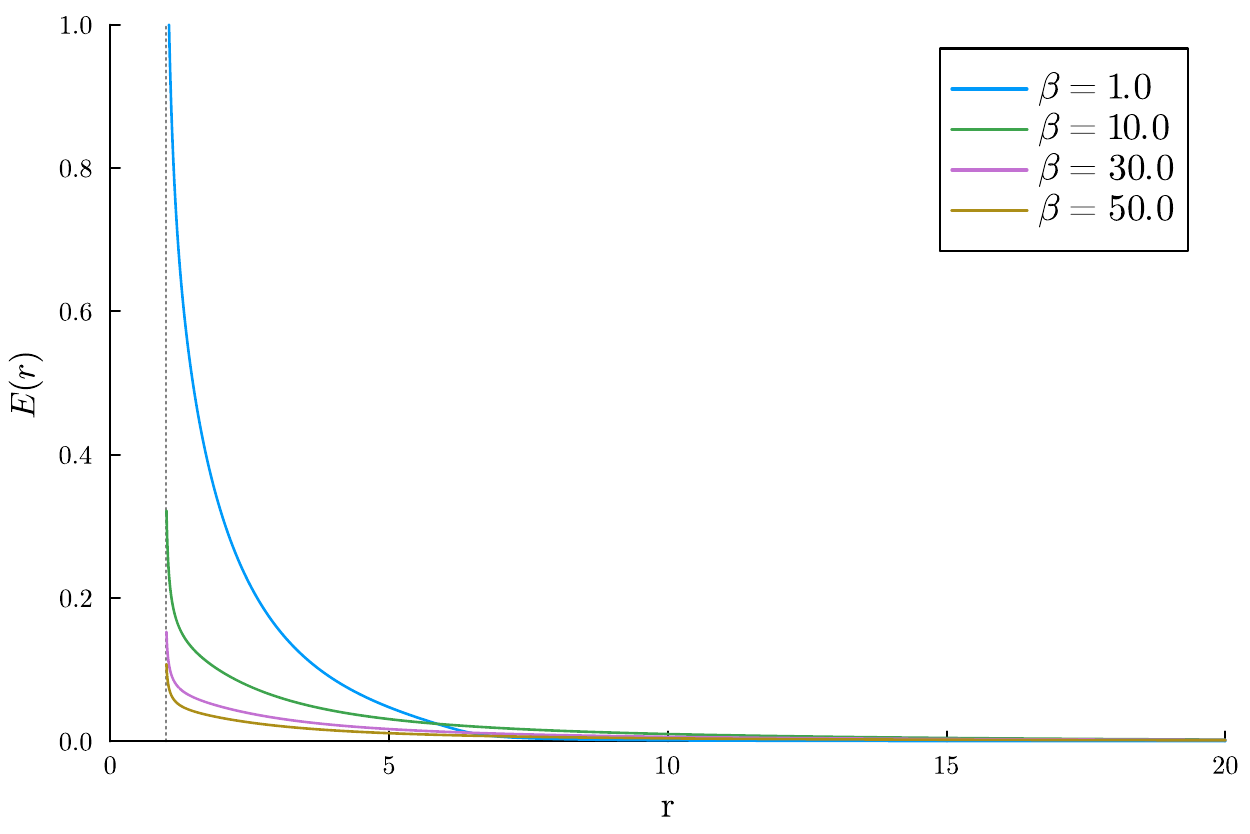}
		\caption{}
		\label{fig:VaryBe-EF_a}
	\end{subfigure}
	\begin{subfigure}{.45\textwidth}
		\centering
		\includegraphics[width=\linewidth]{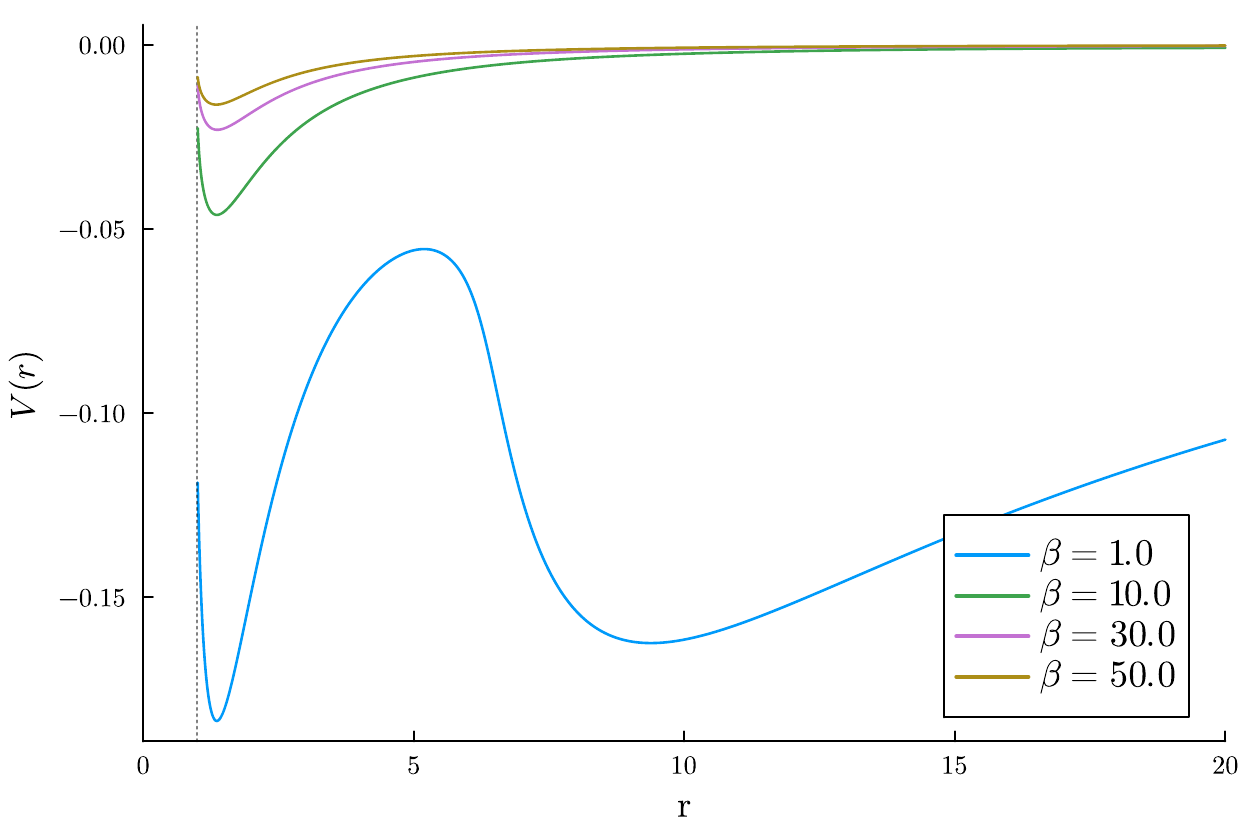}
		\caption{}
		\label{fig:VaryBe-EF_b}
	\end{subfigure}
	\caption{Results for $E(r)$ and $V(r)$ while varying $\beta$ for $Q=0.5$, $\alpha=0.03$, and $N=1.0$.}
	\label{fig:VaryBe-EF}
\end{figure}

Looking at Figure \ref{fig:VaryBe-EF_a}, we note that increasing $\beta$ suppresses the electric field and its potential, the same behaviour seen for $\phi$. Importantly, we see that our $\beta=1$ value in \ref{fig:VaryBe-EF_b} is unstable and demonstrates non-physical behaviour. This acts as a verification for our choice to use larger values of $\beta$ for the other results.

\subsection{Varying N}

\begin{figure}[h]
	\captionsetup[subfigure]{justification=centering}
	\begin{subfigure}{.45\textwidth}
		\centering
		\includegraphics[width=\linewidth]{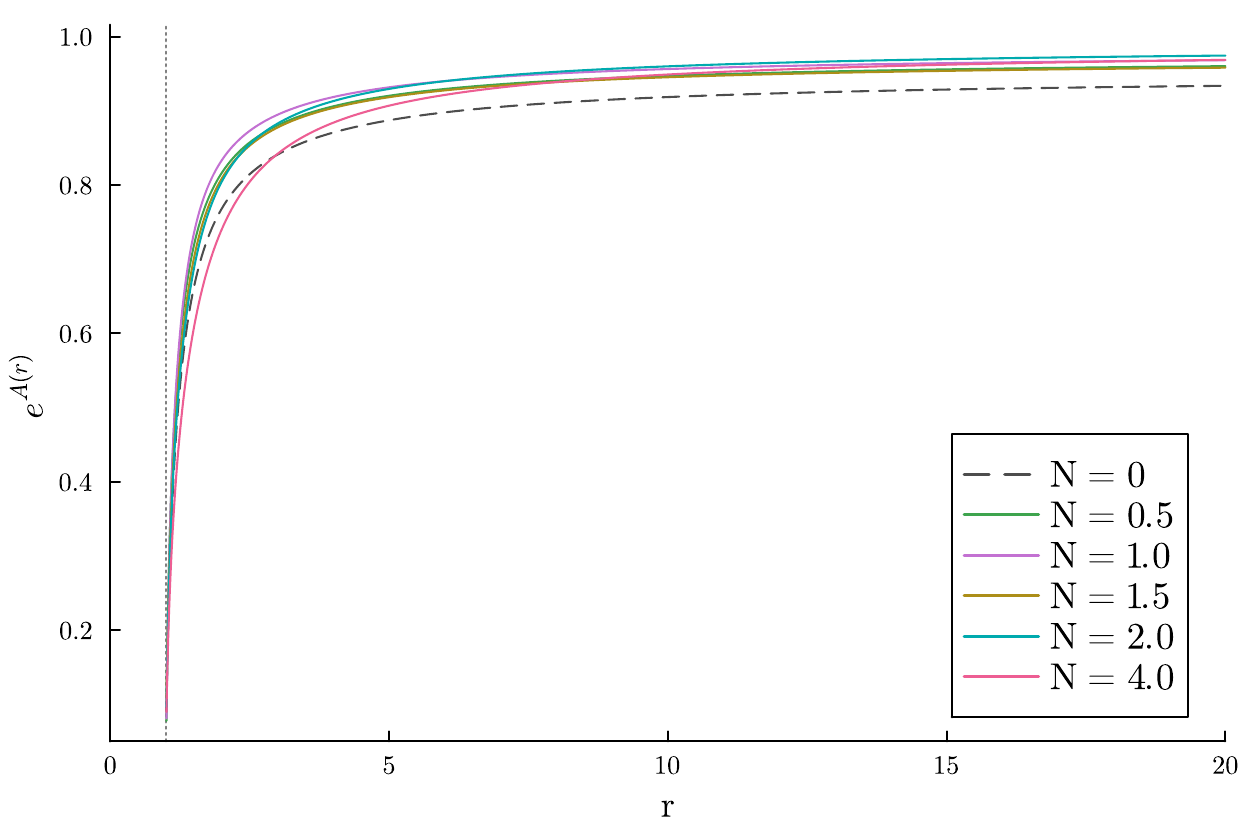}
		\caption{}
		\label{fig:VaryN_a}
	\end{subfigure}
	\begin{subfigure}{.45\textwidth}
		\centering
		\includegraphics[width=\linewidth]{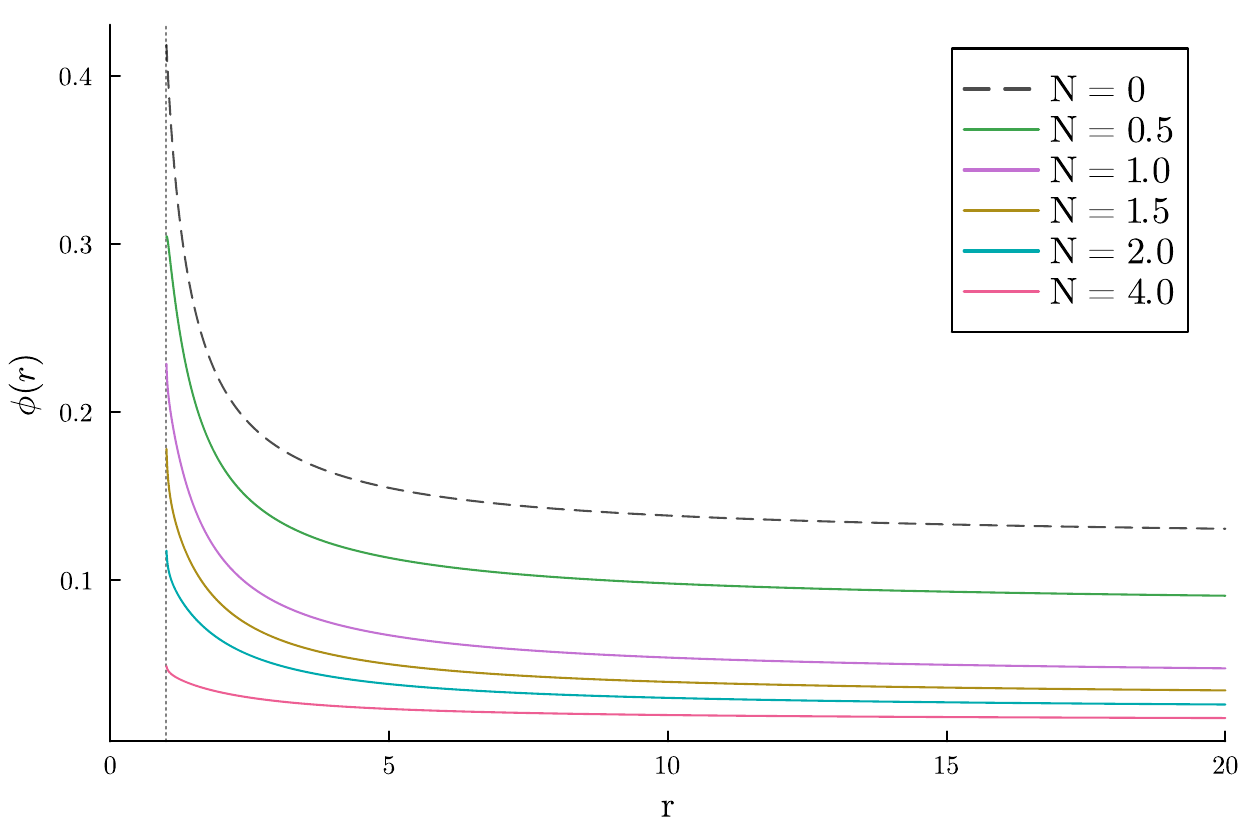}
		\caption{}
		\label{fig:VaryN_b}
	\end{subfigure}
	\caption{Results for $e^{A(r)}$ and $\phi(r)$ while varying $N$ for $Q=0.5$, $\alpha=0.05$, $\beta=30.0$, and $N=1.0$.}
	\label{fig:VaryN}
\end{figure}

Next we look at the effect that the magnitude of the NUT charge $N$ has on our solutions. In Figure \ref{fig:VaryN_a}, we find that $N$ has a minimal impact on the metric time component ${\rm e}^A$ and only diverges closer to the horizon for $N=4$. Overall we find that the presence of $N$ suppresses the magnitude of ${\rm e}^A$ compared to $N=0$ given as a dashed line. This relationship continues for the scalar field as seen in Figure \ref{fig:VaryN_b} but with far clearer scaling with increasing $N$. 
\begin{figure}[h]
	\captionsetup[subfigure]{justification=centering}
	\begin{subfigure}{.45\textwidth}
		\centering
		\includegraphics[width=\linewidth]{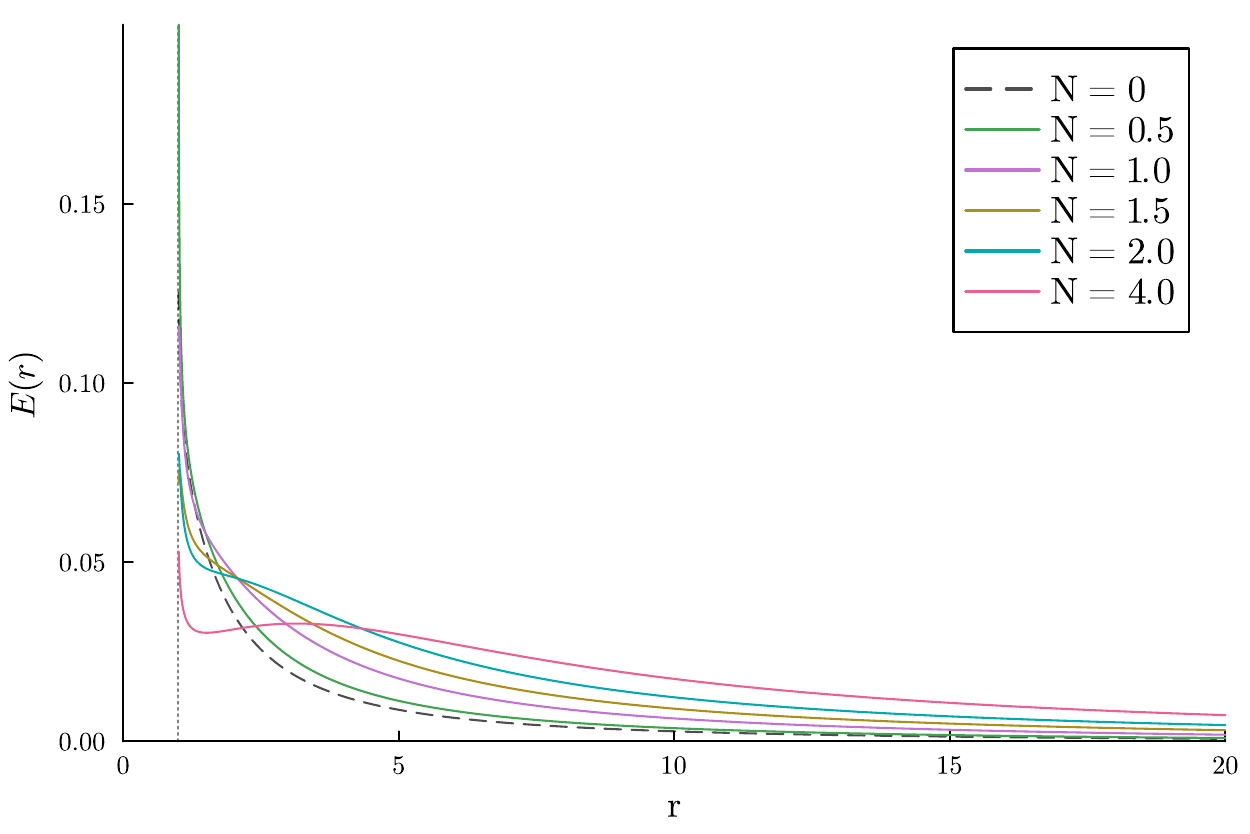}
		\caption{}
		\label{fig:VaryN-EF_a}
	\end{subfigure}
	\begin{subfigure}{.45\textwidth}
		\centering
		\includegraphics[width=\linewidth]{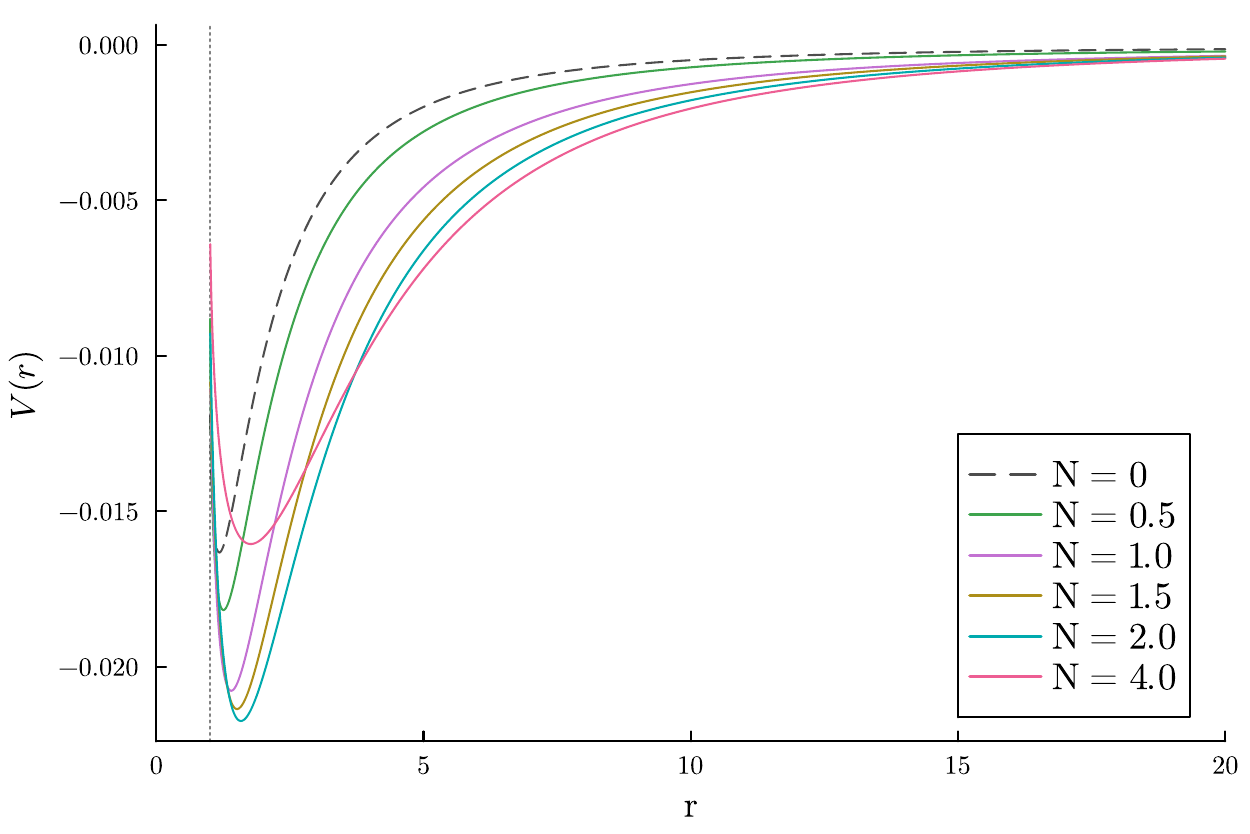}
		\caption{}
		\label{fig:VaryN-EF_b}
	\end{subfigure}
	\caption{Results for $E(r)$ and $V(r)$ while varying $N$ for $Q=0.5$, $\alpha=0.05$, $\beta=30.0$, and $N=1.0$.}
	\label{fig:VaryN-EF}
\end{figure}

There is noteworthy behaviour in the electric field given in Figure \ref{fig:VaryN-EF}, where we observe that a phase change occurs relative to the varying of $N$ at $r\approx2.1$. Larger $N$ are initially weaker near the horizon but the electric field is conversely stronger than smaller $N$ solutions past this inflection point out into the spacetime. Otherwise we find the presence of $N$ initially weakens the electric field near the horizon but produces an overall stronger $E$ field. This behaviour of $E$ may hint at an observable distinction for theories of this kind in the context of gravitomagnetic spacetimes.

The behaviour of $V$ in Figure \ref{fig:VaryN-EF_b} sees the NUT charge contributing to a stronger potential function overall compared to the $N=0$ case. The only solution that behaves differently is $N=4$ which might suggest a constraint on $N$ for this set of parameters.

We attempted to take additional care to ensure this observed behaviour in $E$ was not the result of the numerical methods used. These solutions satisfy all equations with error below $\epsilon = 10^{\textcolor{black}{-8}}$ and we were able to reproduce these results with other implicit Runge-Kutta methods such as \texttt{radau}.

\subsection{Varying N with Q}

\begin{figure}[h]
	\captionsetup[subfigure]{justification=centering}
	\begin{subfigure}{.45\textwidth}
		\centering
		\includegraphics[width=\linewidth]{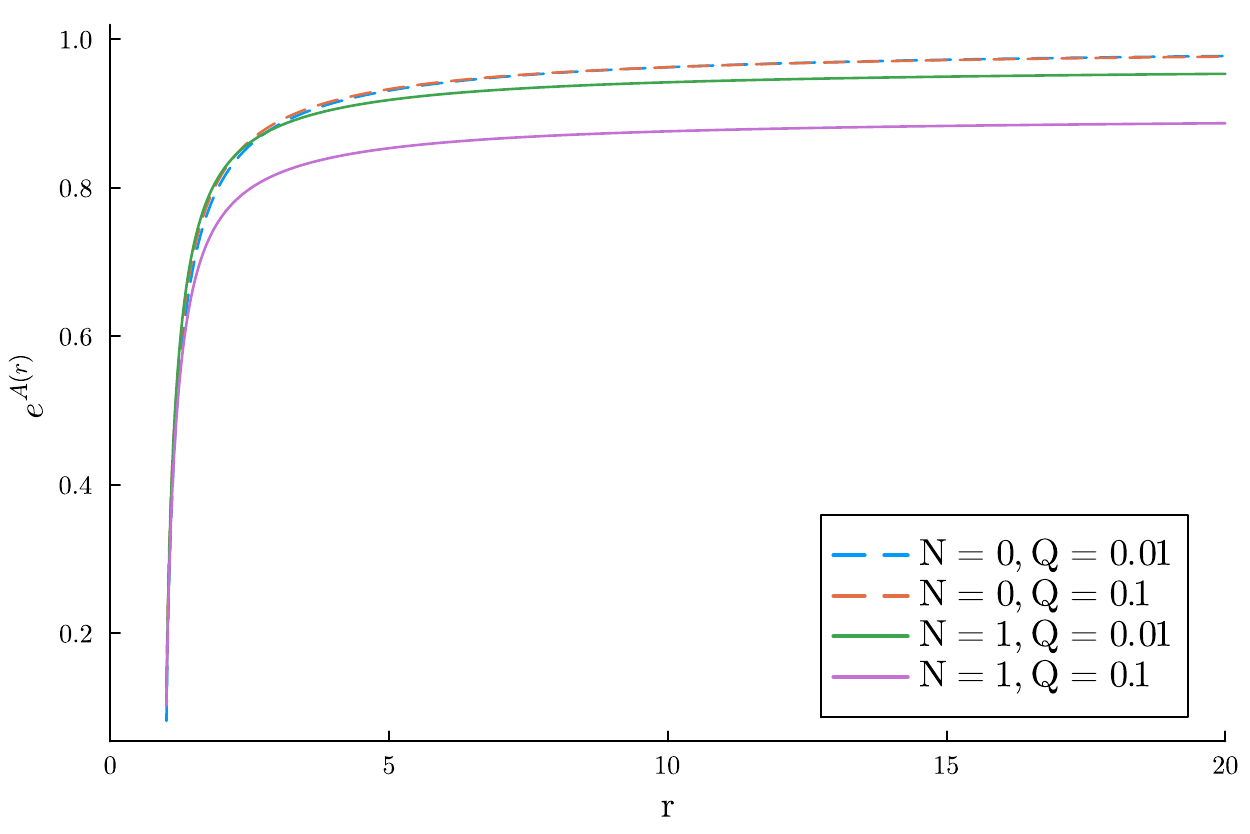}
		\caption{}
		\label{fig:VaryNQ_a}
	\end{subfigure}
	\begin{subfigure}{.45\textwidth}
		\centering
		\includegraphics[width=\linewidth]{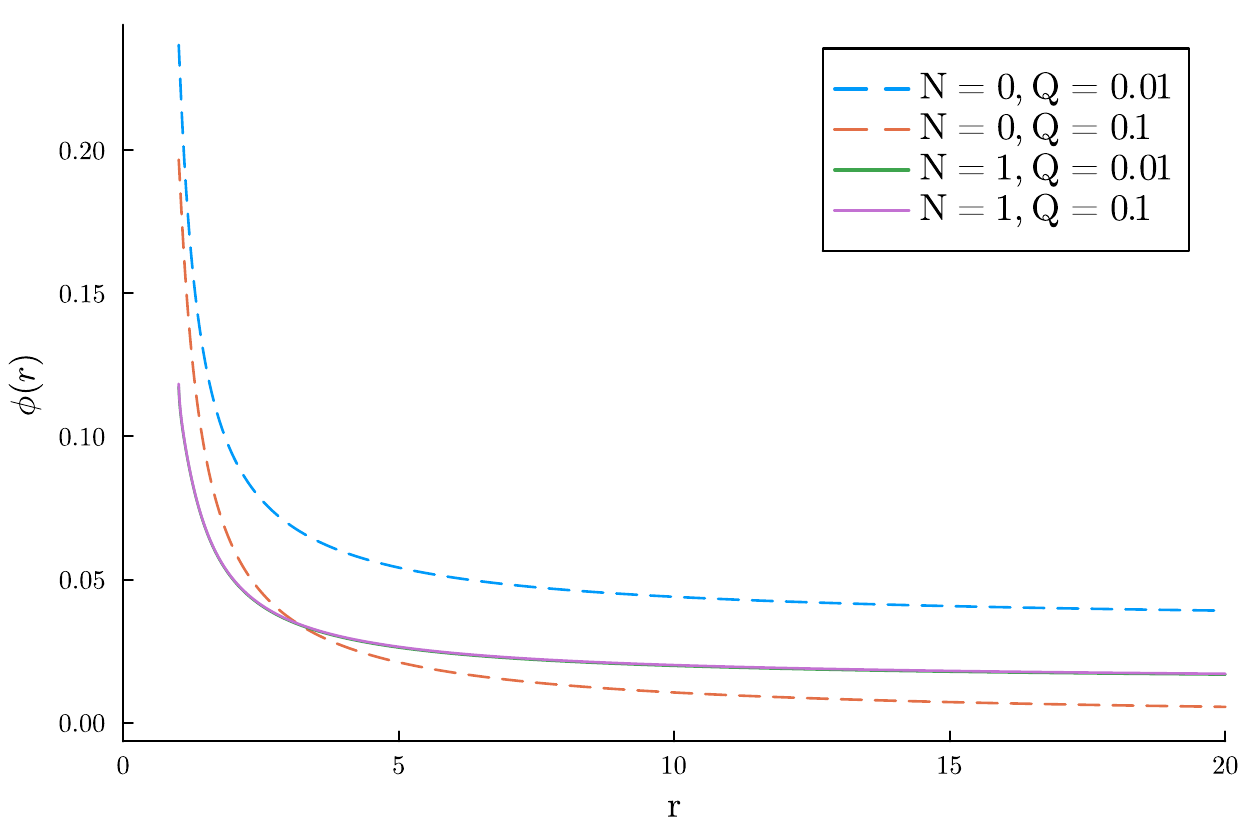}
		\caption{}
		\label{fig:VaryNQ_b}
	\end{subfigure}
	\caption{Results for $e^{A(r)}$ and $\phi(r)$ while varying $N$ with $Q$ where $\alpha=0.05$ and $\beta=0.05$}
	\label{fig:VaryNQ}
\end{figure}

Looking at the relationship between $N$ and $Q$, we find in Figure \ref{fig:VaryNQ_a} that $N$ amplifies the effect that $Q$ has on the time component of the metric. Small changes in $Q$ have little effect if $N=0$ but the magnitude of ${\rm e}^A$ increases substantially with $Q$ when $N=1$. We see the reverse behaviour with the scalar field in Figure \ref{fig:VaryNQ_b}. Without $N$ present, $Q$ will strongly suppress $\phi$ but it has minimal impact when $N=1$. This suggests that the presence of a NUT charge will reverse the influence of $Q$ on the time component and scalar field.

\begin{figure}[h]
	\captionsetup[subfigure]{justification=centering}
	\begin{subfigure}{.45\textwidth}
		\centering
		\includegraphics[width=\linewidth]{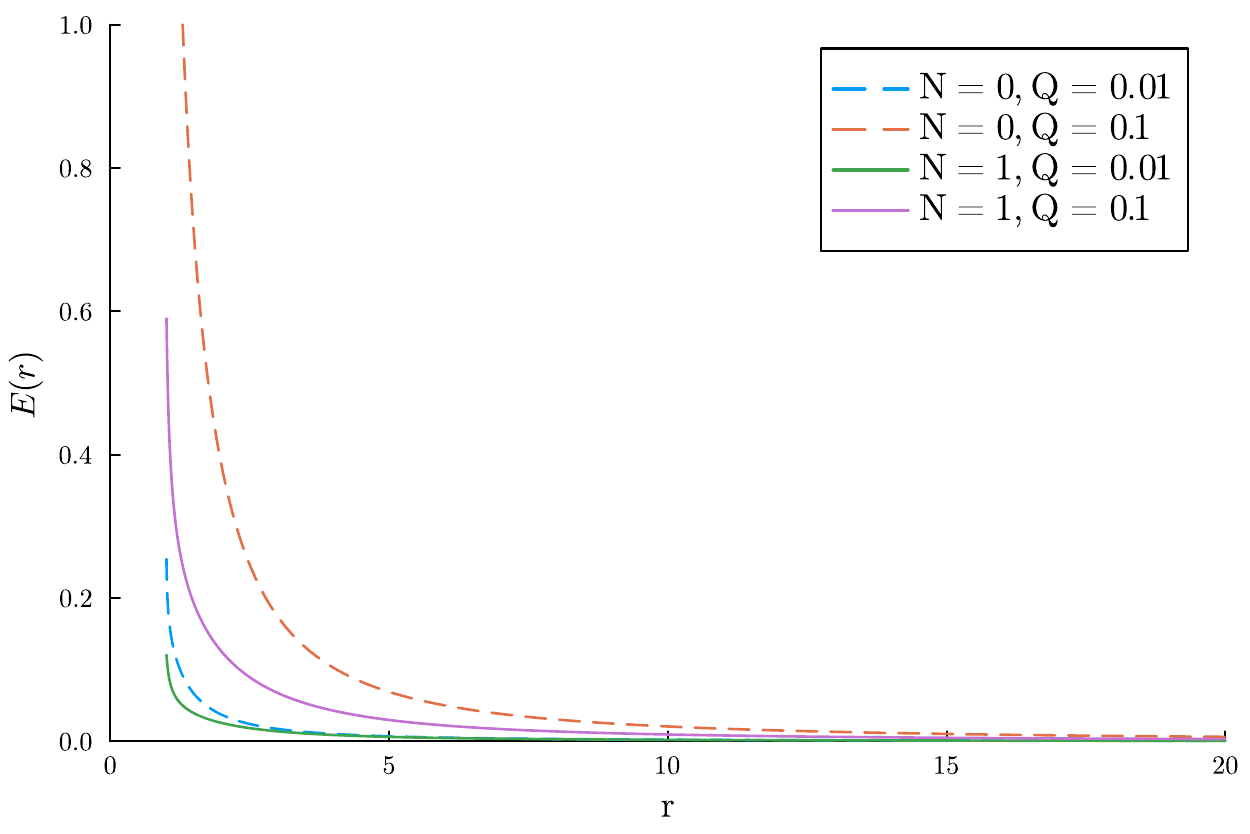}
		\caption{}
		\label{fig:VaryNQ-EF_a}
	\end{subfigure}
	\begin{subfigure}{.45\textwidth}
		\centering
		\includegraphics[width=\linewidth]{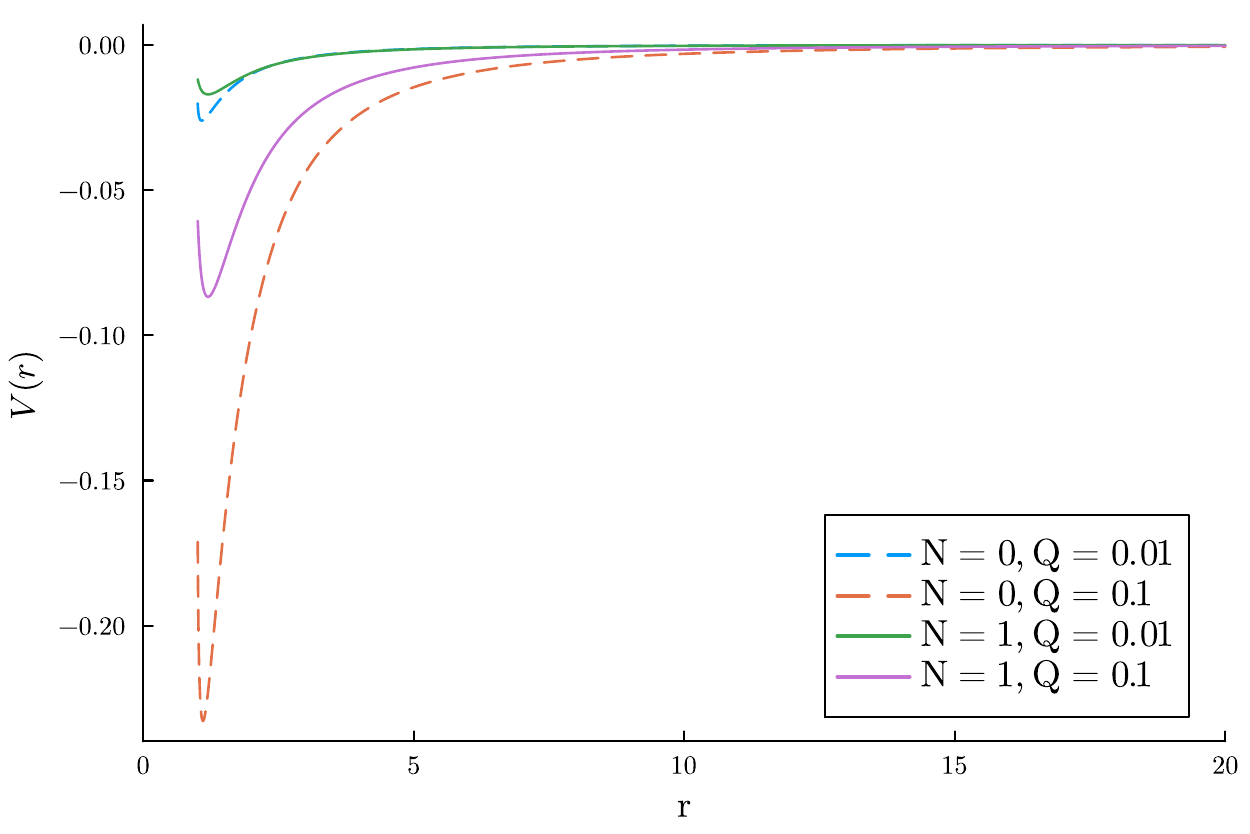}
		\caption{}
		\label{fig:VaryNQ-EF_b}
	\end{subfigure}
	\caption{Results for $E(r)$ and $V(r)$ while varying $N$ with $Q$ where $\alpha=0.05$ and $\beta=0.05$.}
	\label{fig:VaryNQ-EF}
\end{figure}

For the electric field, we see in Figure \ref{fig:VaryNQ-EF} that the NUT charge reduces the effect that smaller $Q$ values have in increasing the magnitude of $E$ and $V$, which is in-line with equations \eqref{eqn:EF} \eqref{eqn:EP}.

\subsection{Varying N with $\alpha$}

\begin{figure}[h]
	\captionsetup[subfigure]{justification=centering}
	\begin{subfigure}{.45\textwidth}
		\centering
		\includegraphics[width=\linewidth]{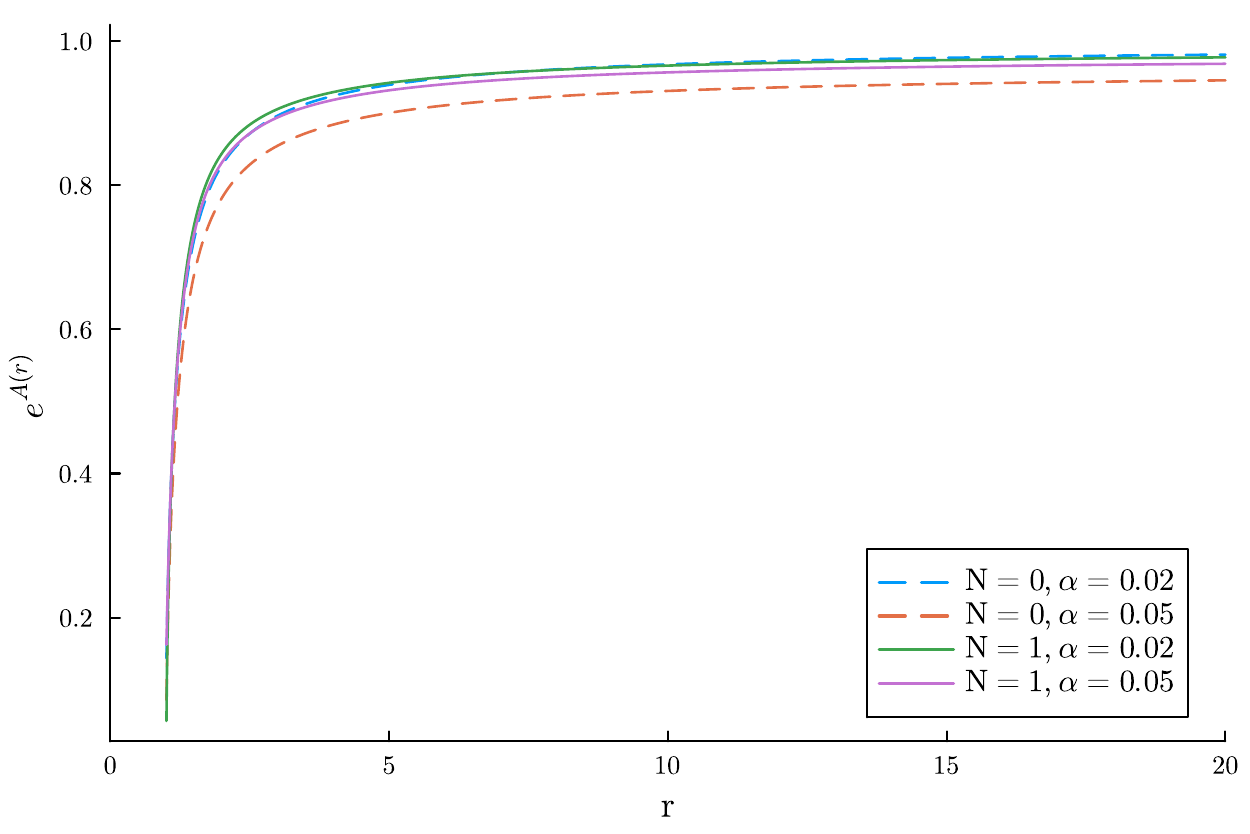}
		\caption{}
		\label{fig:VaryNA_a}
	\end{subfigure}
	\begin{subfigure}{.45\textwidth}
		\centering
		\includegraphics[width=\linewidth]{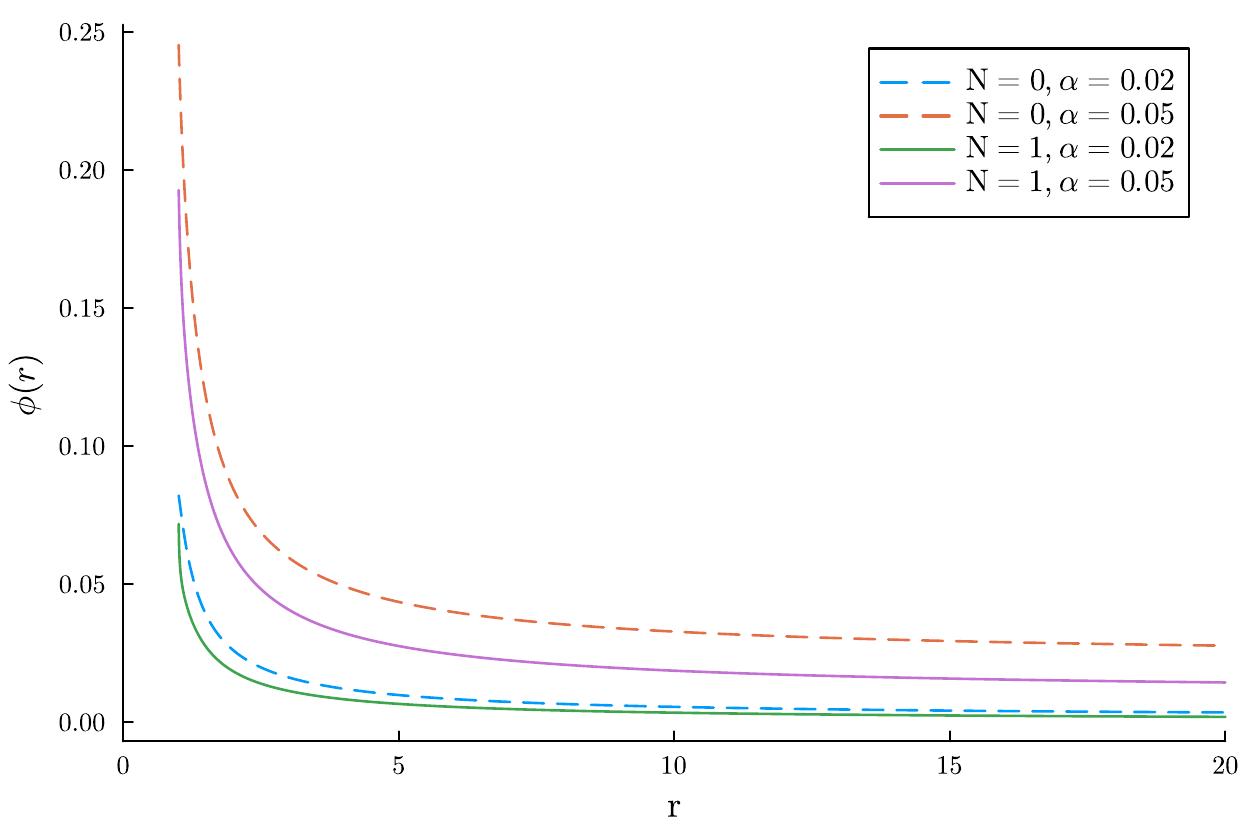}
		\caption{}
		\label{fig:VaryNA_b}
	\end{subfigure}
	\caption{Results for $e^{A(r)}$ and $\phi(r)$ while varying $N$ with $\alpha$ for $Q=0.05$ and $\beta=30.0$.}
	\label{fig:VaryNA}
\end{figure}

In Figure \ref{fig:VaryNA}, we see the same behaviour when we vary $N$ that was observed in Figure \ref{fig:VaryN}. Increases in $N$ result in reduced magnitudes of ${\rm e}^A$ and the scalar field $\phi$. In the presence of a NUT charge, changes in $\alpha$ result in less separation of the solutions for ${\rm e}^A$. The scalar field shows only a minimal dependence between $N$ and $\alpha$, with the largest effect coming from the change in $\alpha$ and with $N$ reducing the solution magnitude overall.
\begin{figure}[h]
	\captionsetup[subfigure]{justification=centering}
	\begin{subfigure}{.45\textwidth}
		\centering
		\includegraphics[width=\linewidth]{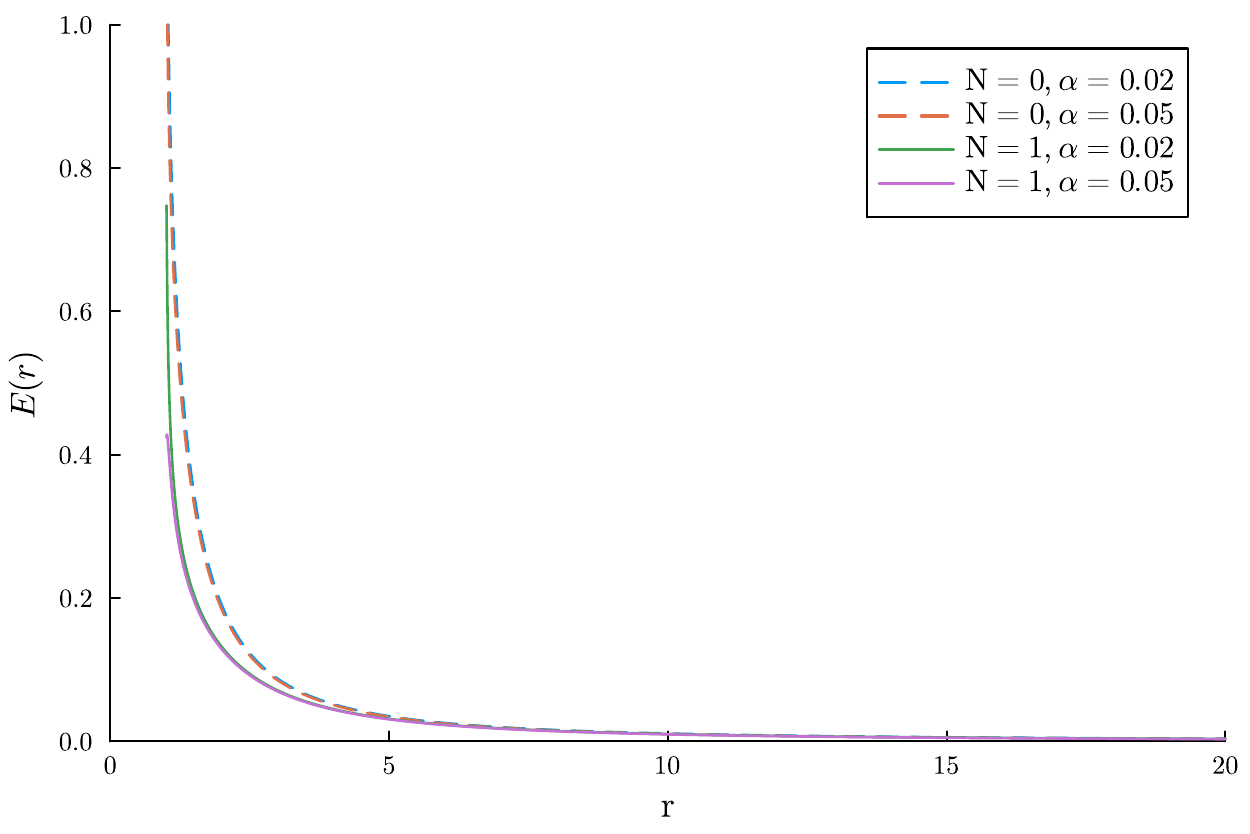}
		\caption{}
		\label{fig:VaryNA-EF_a}
	\end{subfigure}
	\begin{subfigure}{.45\textwidth}
		\centering
		\includegraphics[width=\linewidth]{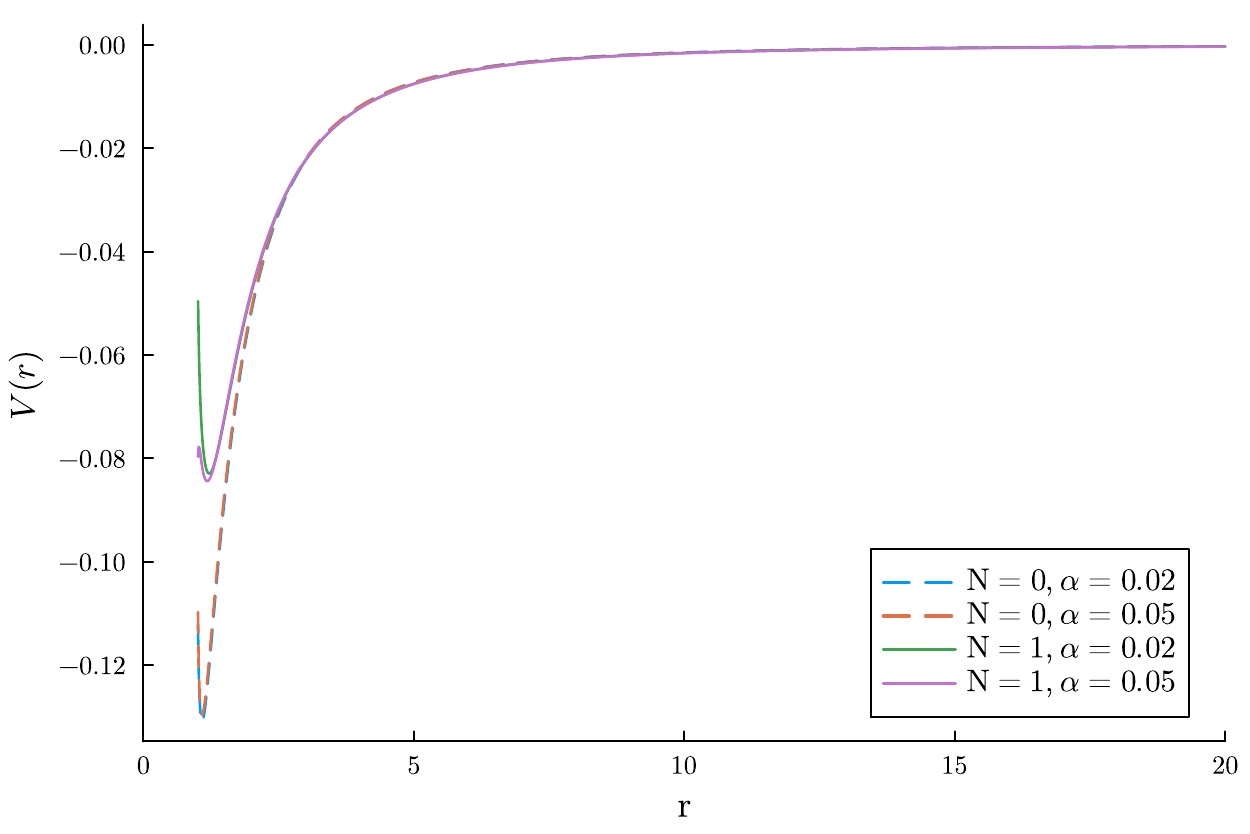}
		\caption{}
		\label{fig:VaryNA-EF_b}
	\end{subfigure}
	\caption{Results for $E(r)$ and $V(r)$ while varying $N$ with $\alpha$ for $Q=0.05$ and $\beta=30.0$.}
	\label{fig:VaryNA-EF}
\end{figure}

In Figure \ref{fig:VaryNA-EF}, we find no noticeable effect of $\alpha$ on the electric field, while the presence of $N$ causes a reduction in the magnitude of $E$ and $V$.

\subsection{Varying N with $\beta$}

\begin{figure}[h]
	\captionsetup[subfigure]{justification=centering}
	\begin{subfigure}{.45\textwidth}
		\centering
		\includegraphics[width=\linewidth]{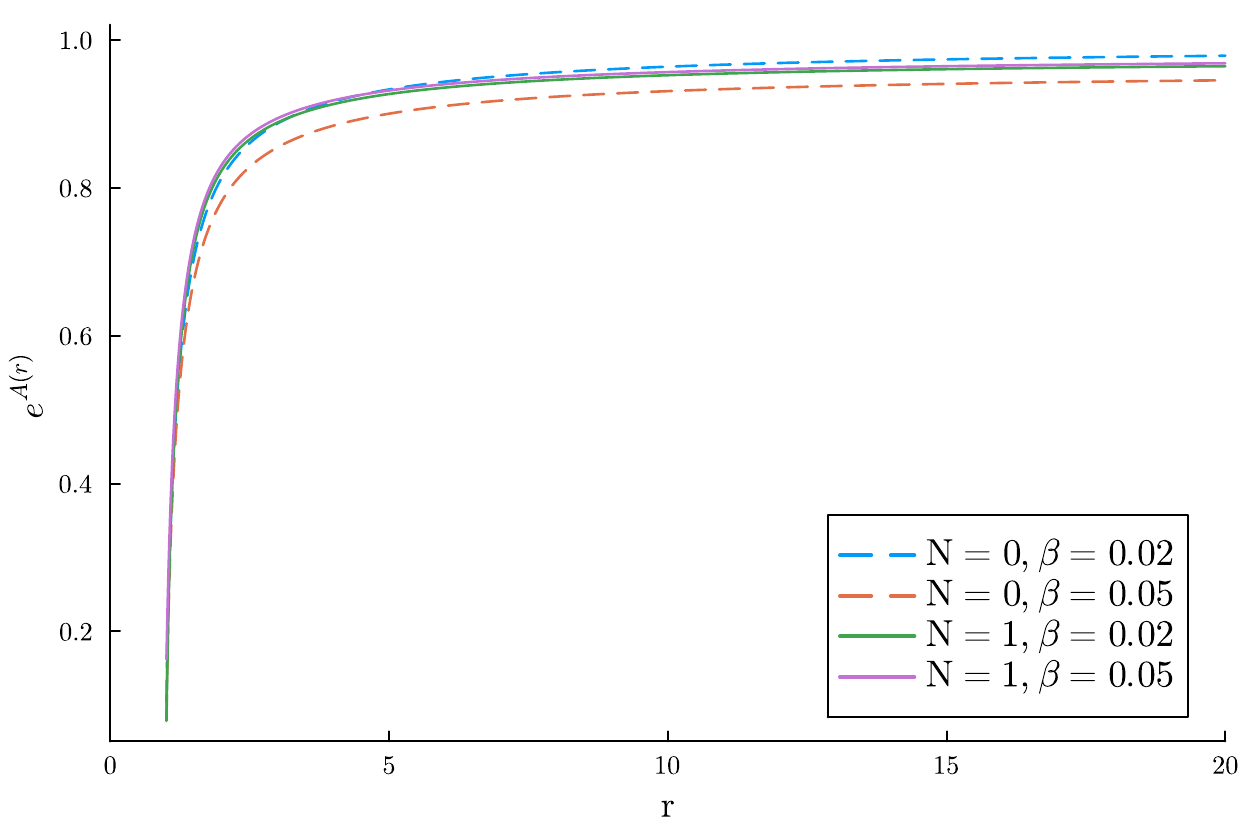}
		\caption{}
		\label{fig:VaryNB_a}
	\end{subfigure}
	\begin{subfigure}{.45\textwidth}
		\centering
		\includegraphics[width=\linewidth]{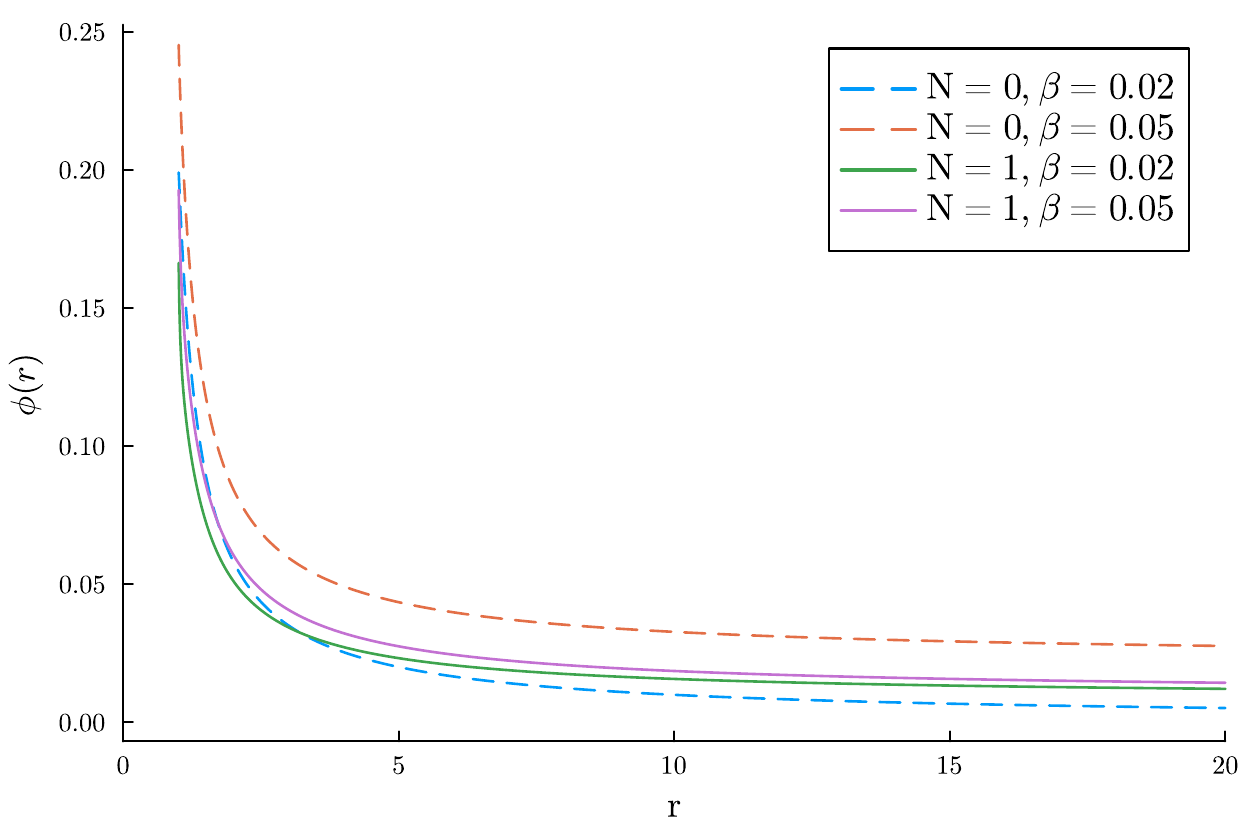}
		\caption{}
		\label{fig:VaryNB_b}
	\end{subfigure}
	\caption{Results for $e^{A(r)}$ and $\phi(r)$ while varying $N$ with $\beta$ for $Q=0.05$ and $\alpha=0.05$}
	\label{fig:VaryNB}
\end{figure}

Finally, examining the relationship between $N$ and $\beta$ we look at Figure \ref{fig:VaryNB_a}. We observe that the presence of the NUT charge minimizes the strong effect of varying $\beta$ on the time component we saw in Figure \ref{fig:VaryBe}, suppressing the increased magnitude of ${\rm e}^e$ in the solutions. Then for the scalar field, we find again that $\beta$ has the largest effect on the solutions and this effect is reduced in the presence of $N$. This behaviour can be produced for larger $Q$ and $\beta$ values as well, but those results offer no additional insights so they were not reported here.
\begin{figure}[h]
	\captionsetup[subfigure]{justification=centering}
	\begin{subfigure}{.45\textwidth}
		\centering
		\includegraphics[width=\linewidth]{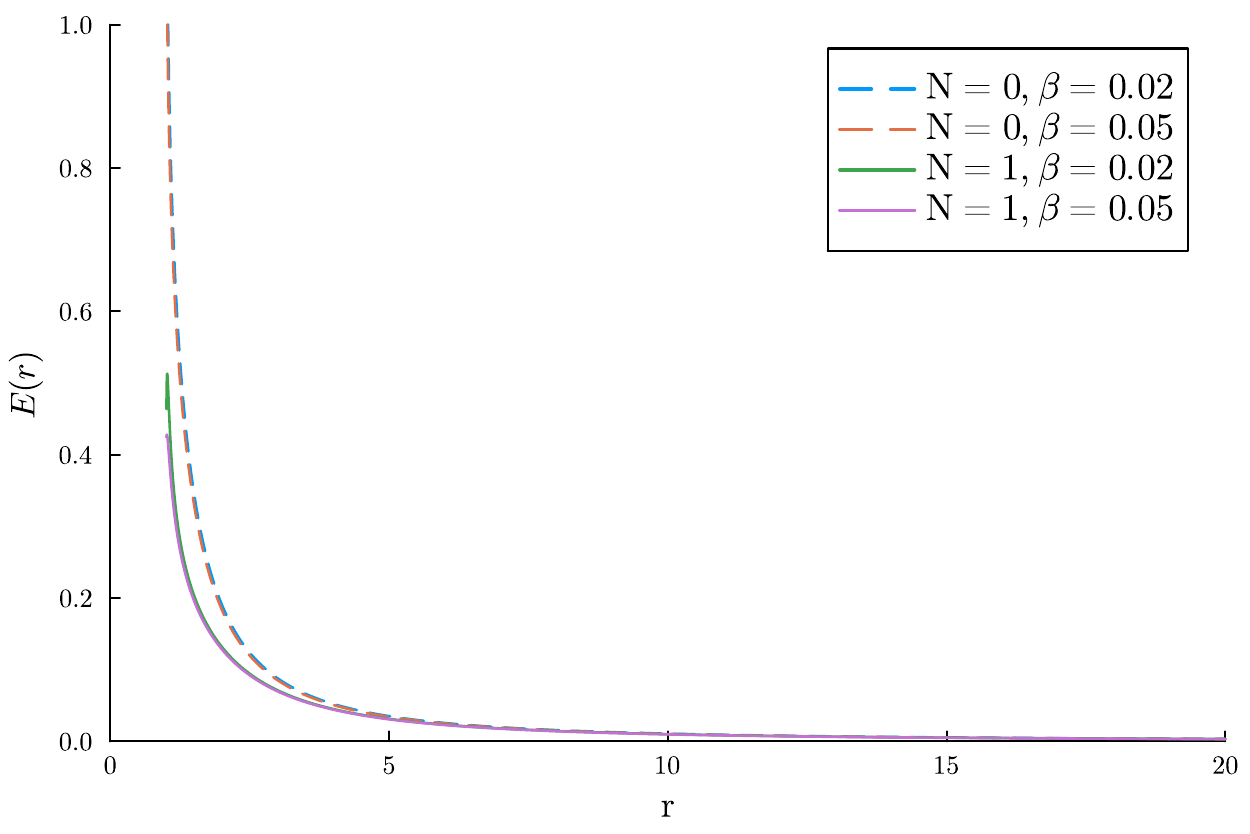}
		\caption{}
		\label{fig:VaryNB-EF_a}
	\end{subfigure}
	\begin{subfigure}{.45\textwidth}
		\centering
		\includegraphics[width=\linewidth]{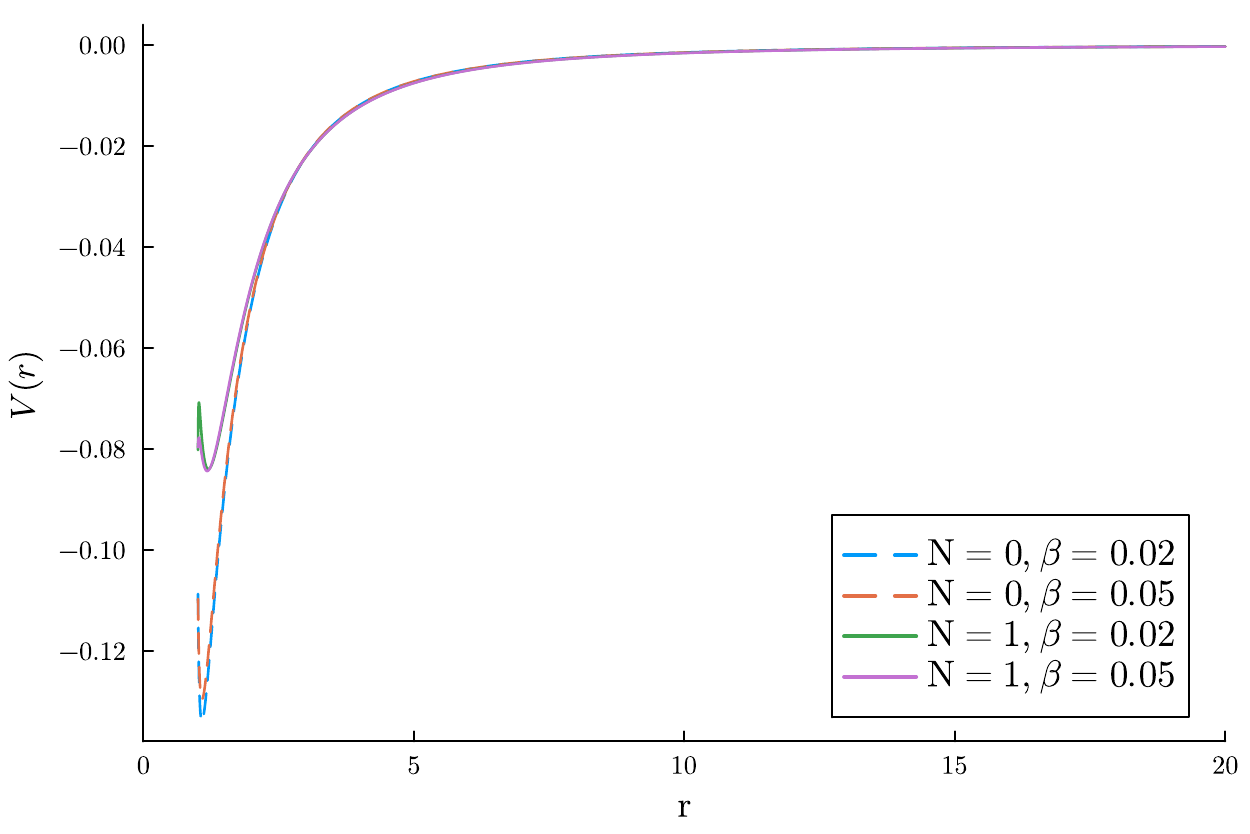}
		\caption{}
		\label{fig:VaryNB-EF_b}
	\end{subfigure}
	\caption{Results for $E(r)$ and $V(r)$ while varying $N$ with $\beta$ for $Q=0.05$ and $\alpha=0.05$.}
	\label{fig:VaryNB-EF}
\end{figure}

In Figure \ref{fig:VaryNB-EF} we find that there is no meaningful dependence between $\beta$ and $N$ regarding the electric field. As with ${\rm e}^A$ and $\phi$, the electric field varies more strongly with a change in $N$ then with $\beta$, with each pair of $\beta$ solutions for a given $N$ being closely grouped together.

\section{Conclusions}\label{sec:conc}

In this article, we construct numerical NUTty solutions to the four-dimensional Einstein-Maxwell-scalar-Gauss-Bonnet gravity where there are two non-minimally couplings between the scalar field and the Gauss-Bonnet invariant and the scalar field to the Maxwell field.  We use an ansatz for the gravitational line element with two unknown metric functions. The ansatz is inspired by the Taub-NUT solutions in the Einstein gravity. We solve numerically all the field equations and discuss their behaviours with respect to the different parameters of the theory. The solutions provide the first known NUTty solutions in the extended theories of gravity, with non-minimal couplings between the scalar field, Gauss-Bonnet term and the Maxwell field. We find that the metric functions are regular  everywhere and the different fields of the theory behave quite steady and consistent in all regions of the spacetime.

We conclude with some comments about extending the results of the article. It is quite possible to find the rotating, as well as the cosmological versions of the solutions, by employing the line element ansatz, similar to the Kerr-NUT-anti-de-Sitter spacetimes. These solutions are of particular interest to establish the black hole holography or study them in the context of AdS/CFT correspondence.  The other possibility is to find numerically the analog of other self-dual geometries such as Eguchi-Hanson, in the four-dimensional Einstein-Maxwell-scalar-Gauss-Bonnet gravity. We also can study the thermodynamics of the NUTty solutions that we leave for a future article.
\appendix
\section{Field Equations} \label{App:A}

The following presents the field equations from calculating equations \eqref{eqn:FE-GR} and \eqref{eqn:FE-scalar} using our metric equation \eqref{eqn:metric-anz}. The formulation presented here is without substitution of ${\rm e}^B$, $B'$, $E$, or $V$. The final forms used with those substitutions are too large to report in a publishable format. The following are the time, radial, the two angular components of $\eqref{eqn:FE-GR}$,
\begin{align}
\begin{split}
	G_{tt} &= \frac{2}{\left(N^{2}+r^{2}\right)^{3}} 
	\bigg[
	\tfrac32\left(N^{4} + N^{2} r^{2}\right) {\rm e}^{2 (A + B)} - 4\left(N^{2} \beta  \,V^{2} \phi + \tfrac18 N^{2} V^{2} - \tfrac18 N^{2}- \tfrac18 r^{2}\right) \\
	& \times \left(N^{2}+r^{2}\right) {\rm e}^{A + 2 B} + 6 N^{2} \alpha  \left(N^{2}+r^{2}\right) \left(\phi' B' -2 \phi'' \right) {\rm e}^{2 A + B} \\
	& - \tfrac{1}{4} \left(N^{2}+r^{2}\right) \left(16 \alpha  \left(N^{2}+r^{2}\right) \phi'' \right. 
	\left. + \left(N^{2}+r^{2}\right)^{2} \phi'^{2}-8 B' \alpha  \left(N^{2}+r^{2}\right) \phi' \right. \\
	&\left. -\left(2 N^{2} r + 2 r^{3}\right) B' +4 N^{2}+2 r^{2} \right) {\rm e}^{A +B} 
	+ 4 r^{2} \alpha  \,{\rm e}^{A} \left(N^{2}+r^{2}\right) \phi'' \\
	&- 6r \left(\left(N^{2} r +r^{3}\right) B' -\tfrac43 N^{2}\right) \alpha  \,{\rm e}^{A} \phi' - E^{2}  \left(\beta  \phi + \tfrac18 \right){\rm e}^{B} \left(N^{2}+r^{2}\right)^{3}
	\bigg] {\rm e}^{-2B},
\end{split}
\label{eqn:DE-Gtt}
\end{align}
\begin{align}
\begin{split}
	G_{rr} & = \frac{2}{\left(N^{2}+r^{2}\right)^{3}}	
	\bigg[ 
	-\tfrac{1}{2}\left(N^{4}+N^{2} r^{2}\right) {\rm e}^{2(A + B)} + 4 \left(N^{2} \beta  \,V^{2} \phi + \tfrac18 N^{2} V^{2}-\tfrac18 N^{2}- \tfrac18 r^{2}\right)\\ 
	& \times \left(N^{2}+r^{2}\right) {\rm e}^{A + 2 B}+6 \left(\left(N^{2}+r^{2}\right) A' - \tfrac43 r\right) \alpha  \,N^{2} \phi' \,{\rm e}^{2A + B} - \left(\left(\tfrac{1}{4} \left(N^{2}+r^{2}\right)^{3} \phi'^{2} \right.\right. \\
	& \left. - 2 A' \alpha  \left(N^{2}+r^{2}\right)^{2} \phi' - \tfrac{1}{2} \left(\left(N^{2}+r^{2}\right)^{2} A' +r \right)\right) r {\rm e}^{A +B} \\
	& \left. + 6 A' \phi' \,r^{2} \alpha  \,{\rm e}^{A} - E^{2} \left(\beta  \phi + \tfrac18\right) {\rm e}^{B} \left(N^{2}+r^{2}\right)^{3}\right)
	\bigg] {\rm e}^{-(A + B)},
\end{split}
\label{eqn:DE-Grr}
\end{align}
\begin{align}
\begin{split}
	G_{\theta\theta} &= 
	\frac{2}{\left(N^{2}+r^{2}\right)}  
	\bigg[
	\tfrac{1}{2} N^{2} {\rm e}^{2(A+B)}  + 2 N^{2} \alpha  \left(\phi' B' - 2\phi'' \right) {\rm e}^{2A + B} - 4 N^{2} \left(\beta  \phi + \tfrac18\right) V^{2} {\rm e}^{A + 2B}\\
	& - \tfrac18 \left(2 \left(N^{2}+r^{2}\right)^{2} A'' - \left(N^{2}+r^{2}\right)^{2} A'^{2} + \left(\left(N^{2}+r^{2}\right) B' -2 r \right) \left(N^{2}+r^{2}\right) A' \right. \\
	& - 2\left(N^{2}+r^{2}\right)^{2} \phi'^{2}  
	\left. + 2\left(N^{2} r +r^{3}\right) B' - N^{2}\right) {\rm e}^{A+B} - 2\,{\rm e}^{A} \phi' r \alpha  \left(N^{2}+r^{2}\right) A'' \\
	& - 2\,{\rm e}^{A} A' r \alpha  \left(N^{2}+r^{2}\right) \phi''
	- {\rm e}^{A} \phi' r \alpha  \left(N^{2}+r^{2}\right)A'^{2} \\
	& + 3\phi' \left(\left(N^{2} r +r^{3}\right) B' - \tfrac{2}{3} N^{2}\right) \alpha  \,{\rm e}^{A} A'
	 - {\rm e}^{B} E^{2} \left(\beta  \phi + \tfrac18 \right) \left(N^{2}+r^{2}\right)^{2}
	\bigg] {\rm e}^{-(A + 2 B)},
\end{split}
\label{eqn:DE-Gth}
\end{align}
\vfill
\begin{align}
\begin{split}
	G_{\phi\phi} & = -\frac{8}{\left(N^{2}+r^{2}\right)^{3}}
	\bigg[
	N^{2} \left(N^{2}+r^{2}\right) \left[  -\tfrac{3}{2}N^{2}  {\rm e}^{3 A + 2 B} \cos^{2}\theta \right. \\
	& \tfrac18  \left(\left(32 N^{2} \beta  \,V^{2} \phi + 4 N^{2} V^{2} - 3 N^{2} - 3 r^{2}\right) {\rm e}^{2(A+B)} \cos^{2}\theta  + \left(N^{2}+r^{2}\right)\right) \\
	& + \tfrac{1}{4} \left\{ \left(12 \alpha \phi'' -6 \alpha  B' \phi' \right) \left(\cos^{2}\theta +\tfrac{1}{3}\right) \left(N^{2}+r^{2}\right)  +\cos^{2}\theta \left(N^{2}+r^{2}\right)^{2} \phi'^{2} \right.\\
	&\left.  -2 \cos^{2}\theta \left( \left(N^{2} +r^{2}\right) B'r -2 N^{2}-r^{2}\right)\right\} {\rm e}^{2 A + B} \\
	& +\sin^{2}\theta \left(\beta  \phi +\tfrac18 \right) V^{2} \left(N^{2}+r^{2}\right) {\rm e}^{A + 2B}
	- 6 N^{2} \cos^{2}\theta \alpha  \left(\phi' B' -2 \phi'' \right) {\rm e}^{3A + B} \\
	&- \tfrac{1}{32 N^2} \left\{2 \left(N^{2}+r^{2}\right)^{2} \sin^{2}\theta A'' + 2 \left(N^{2}+r^{2}\right)^{2} \sin^{2}\theta \phi'^{2} + \left(N^{2}+r^{2}\right)^{2} \sin^{2}\theta  A'^{2} \right.\\
	&- \left(\left(N^{2}+r^{2}\right) B' -2 r \right) \sin^{2}\theta \left(N^{2}+r^{2}\right) A' - 2 \sin^{2}\theta \,r \left(N^{2}+r^{2}\right) B' \\
	& \left. - \left(\cos^{2}\theta \left(\beta  \phi + \tfrac18 \right) \left(N^{2}+r^{2}\right) E^{2} - \tfrac18 \sin^{2}\theta\right) N^{2}\right\}
	\left(N^{2}+r^{2}\right) {\rm e}^{A +B}\\
	&\left. - \tfrac{1}{2} \left(8 N^{2} \cos^{2}\theta r \,{\rm e}^{2 A} - {\rm e}^{A} A' \sin^{2}\theta \left(N^{2}+r^{2}\right)^{2}\right) r \alpha  \phi'' \right.\\
	& \left. + \tfrac{1}{2 N^2}  \sin^{2}\theta \alpha  r \left(N^{2}+r^{2}\right)^{2} {\rm e}^{A} \phi' A'' \right] + 6 r  \alpha  \cos^{2}\theta N^{2} \left(r\left(N^{2} + r^{2}\right) B' -\tfrac{4}{3} N^{2}\right) {\rm e}^{2 A} \phi' \\
	&+ \tfrac{1}{4} \left( \left(r\left(N^{2} +r^{2}\right) A' -3r \left(N^{2} + r^{2}\right) B' +2 N^{2}\right) {\rm e}^{A} \alpha \phi' A' \right. \\ 
	& \left. +{\rm e}^{B} E^{2} \left(\beta  \phi + \tfrac18 \right) \left(N^{2}+r^{2}\right)^{2}\right)
	\sin^{2}\theta \left(N^{2}+r^{2}\right)^{2}
	\bigg] {\rm e}^{-(A + 2B)}.	
\end{split}
\label{eqn:DE-Gpp}
\end{align}
We also include the off-diagonal component of \eqref{eqn:FE-GR},
\begin{align}
\begin{split}
	G_{t\phi} &= \frac{4N}{\left(N^{2}+r^{2}\right)^{2}}
	\bigg[
	\tfrac{3 N^2 }{2} {\rm e}^{A+B}
	- \tfrac{1}{2} \left(8 N^{2} \beta  \,V^{2} \phi + N^{2} V^{2} - N^{2} - r^{2}\right) {\rm e}^{A + 2B} \\
	&+ 6 N^{2} \alpha \left(\phi' B' -2 \phi'' \right) {\rm e}^{B +2 A} 
	- \tfrac{1}{4}\left\{
	 \left(16 \alpha \phi'' + \left(N^{2}+r^{2}\right) \phi'^{2} 
	 - 8 B' \alpha \phi' - 2r B' \right) \right.\\
	& \left. + \tfrac{4 N^{2} + 2 r^{2}}{N^2 + r^2} \right\} {\rm e}^{A +B} 
	+ 4 r^{2} \alpha  \,{\rm e}^{A} \phi'' 
	- 6 r \left(r B' -\tfrac{4N^{2}}{3 \left(N^2 + r^2\right)} \right) \alpha  \,{\rm e}^{A} \phi' \\
	&- E^{2} \,{\rm e}^{B} \left(\beta  \phi + \tfrac18 \right) \left(N^{2}+r^{2}\right)^{2}
	\bigg] {\rm e}^{-2 B} \cos\theta. 
\end{split}
\label{eqn:DE-Gtp}
\end{align}

Finally, we present the single differential equation calculated for the scalar field using equation \eqref{eqn:FE-scalar},
\begin{align}
\begin{split}
	0 &= 
	\frac{\alpha}{2\, \left(N^2 + r^2\right)^4 {{\rm e}^{2B}}} 
	\bigg[2\, \left(4 \,r^2 - \left( 3 {{\rm e}^{A}} N^2 + 4\left(N^2 + r^2 \right) \right) {{\rm e}^{B}}  \right)  \left(N^2+r^2 \right)^2 A''\\
	& + \left( 4 \,r^2 - \left( 9{{\rm e}^{A}}N^2 + 4\,N^2 + 4 \,r^2 \right) {{\rm e}^{B}} \right)  \left( N^2+r^2 \right)^2 A'^2 
	+ 3\,\left( N^2+r^2\right)  \\
	& \times \left\{\left(\left( {{\rm e}^{A}}N^2 + \tfrac43 \,N^2 + \tfrac43 \,r^2 \right) {{\rm e}^{B}} - 4\,r^2 \right) \left( N^2+r^2 \right) B'
	+ 8\,N^2 \left( {{\rm e}^{A + B}} + \tfrac23 \right) r \right\} A'  \\
	&- 4\,N^2{{\rm e}^{A}}{{\rm e}^{B}} \left( r \left(N^2 + r^2\right) B' - 2\,N^2 + 6\,r^2 \right)  
	\bigg] \\
	&+ \left(\phi'' + \tfrac12(A' - B')\frac { 2\,r \phi' }{N^2 + r^2} \right){{\rm e}^{-B}}
	+ \left( \frac{2 \, E^2}{{{\rm e}^{A + B}}} - \frac{8\,N^2}{ \left( N^2+r^2 \right)^2} V^2 \right) \beta. 
\end{split}
\label{eqn:DE-Scalar}
\end{align}

\vspace{1cm} 
\bigskip

{\Large Acknowledgments}\newline
This work was supported by the Natural Sciences and Engineering Research Council of Canada.\newline

\end{document}